\documentclass[twocolumn]{aastex631}
\usepackage{type1cm} 
\usepackage[utf8]{inputenc}
\usepackage{makeidx}         
\usepackage{graphicx}  
\usepackage[bottom]{footmisc}
\usepackage{xspace}
\usepackage{newtxtext}       %
\usepackage{newtxmath}       
\usepackage{comment}
\usepackage{amsmath}

\newcommand{\be}{\begin{equation}}
\newcommand{\ee}{\end{equation}}

\newcommand{\bea}{\begin{eqnarray}}
\newcommand{\eea}{\end{eqnarray}}

\usepackage{soul}

\shorttitle{Universal relation of accreting magnetic stars in GRBs}
\shortauthors{Dall'Osso et al.}

\begin{document}

\title{MAGNETAR CENTRAL ENGINES IN GAMMA-RAY BURSTS FOLLOW THE UNIVERSAL RELATION OF ACCRETING MAGNETIC STARS}
\correspondingauthor{simoneda@roma1.infn.it}
\author[0000-0003-4366-8265]{Simone Dall'Osso}
\affil{Istituto Nazionale di Fisica Nucleare - Roma 1, Piazzale Aldo Moro 2, 00185, Roma}
\author[0000-0003-1055-7980]{Giulia Stratta}
\affil{Istituto Nazionale di Fisica Nucleare - Roma 1, Piazzale Aldo Moro 2, 00185, Roma}
\affil{Institut f\"{u}r Theoretische Physik, Goethe Universit\"{a}t, Max-von-Laue-Str. 1, 60438 Frankfurt am Main, Germany}
\affil{Istituto di Astrofisica e Planetologia Spaziali, via Fosso del Cavaliere 100, I-00133 Roma, Italy}
\affil{INAF - Osservatorio Astornomico di Bologna, viale P.Gobetti, Bologna}
\author[0000-0002-3635-5677]{Rosalba Perna}
\affil{Departemnt of Physics \& Astronomy, Stony Brook University, Stony Brook, NY, 11794, USA}
\affil{Center for Computational Astrophysics, Flatiron Institute, New York, NY 10010, USA}
\author[0000-0003-0869-7180]{Giovanni De Cesare}
\affil{INAF - Osservatorio Astornomico di Bologna, viale P.Gobetti, Bologna}
\author[0000-0002-0018-1687]{Luigi Stella}
\affil{INAF - Osservatorio Astornomico di Roma, via Frascati 33, 00078, Monte Porzio Catone, Roma}

\begin{abstract}
Gamma-ray bursts (GRBs), both long and short, are explosive events whose inner engine is generally expected to be a black hole or a
highly magnetic neutron star (magnetar) accreting high density matter.~Recognizing~the nature of GRB central engines, and in particular the formation of neutron stars (NSs), is of high astrophysical significance.~A possible signature of NSs in GRBs is the presence of a plateau in the early X-ray afterglow.~Here we carefully select a subset of long and short GRBs with a clear plateau, and look for an additional NS signature in their prompt emission, namely a transition between accretion and propeller in analogy with accreting, magnetic compact objects in other astrophysical sources.~We estimate from the prompt emission the minimum accretion luminosity below which the propeller mechanism sets in, and the NS magnetic field and spin period from the plateau.~We demonstrate that these three quantities obey the same universal relation in GRBs as in other accreting compact objects switching from accretion to propeller.~This relation provides also an estimate of the radiative efficiency of GRBs, which we find to
be several times lower than radiatively efficient accretion in X-ray binaries and in agreement with theoretical expectations.~These results provide additional support to the idea that at least some GRBs are powered by magnetars surrounded by~an~accretion~disc.
\end{abstract}
\section{Introduction}
After several decades of observations, modeling and theoretical investigations,
Gamma-Ray Bursts (GRBs), the most energetic explosions in the Universe,
still leave open questions.~The two generally distinctive classes of long and short GRBs have been securely associated with the collapse of massive stars for the former \citep{hjorth03,stanek03}, and with the merger of two neutron stars (NSs) for the latter \citep{Abbott2017a}.~However, the fate of the remnant compact object left behind after the stellar collapse or the binary NS merger still remains to be deciphered, a problem with implications also for the equation of state of ultra-dense matter \citep{lattimer01}  and the generation of gravitational wave signals (e.g.~\citealt{dallo15,dallo18}).

If the remnant compact object is a NS, it is likely to be highly magnetized.~In the case of formation in the collapse of a massive star, this is because of the fast rotation likely necessary to  drive a long GRB \citep{macfadyen99,heger03}, which facilitates magnetic field amplification due to dynamo action \citep{thompson95,raynaud20,aloy21,ober21,guilet22}.~In the case of NS-NS mergers, magnetic field amplification via dynamo has been demonstrated in general relativistic magnetohydrodynamic simulations of the merger \citep{giacomazzo13,zrake13,giacomazzo15,palenzuela15,kawamura16,ciolfi17,radice17,aguilera20,palenzuela22,aguilera22}. 

The possibility that the 
inner engine driving the GRB emission may be a fast-spinning and highly magnetized NS rather than a black hole has received considerable attention in recent years
(\citealt{thompson04,metzger07,bucciantini07,bucciantini12, dallosso11,Rowlinson13,beniamini17,metzger18}; see however, e.g.,
~\citealt{kumar08,siegel14, beniamini17a, oganesyan20, beniamini20}).~If~GRBs are indeed powered by a magnetar, there are observational signatures that may point to their presence (e.g., \citealt{dai98,zhang01}).~In particular, since the early-time spindown luminosity of a NS has a plateau, this behaviour may be reflected in the initial plateau of the X-ray afterglow, assuming that its luminosity tracks the energy injection from the NS. By fitting the observed light curves, in particular the plateau luminosity and duration, these models provide estimates of the magnetic fields and spin periods of the putative magnetars (e.g.~\citealt{dallosso11,Rowlinson13,gompertz14,lu14,gibson17}).~The resulting $P-$ and $B-$values in the population are found to be correlated in a way that matches the well-established ``spin-up line" of radio pulsars and magnetic accreting NSs in galactic 
X-ray binaries (e.g.~\citealt{bhatta91,pan13}), once re-scaled~for~the much larger mass accretion rates of GRBs~\citep{stratta18}.

In this work we re-examine the impact of a magnetar central engine by considering both prompt and afterglow emissions jointly. Similarly to previous studies (e.g.~\citealt{bernardini13, stratta18}), we assume the prompt is powered by accretion energy while the afterglow plateau by the injection of the NS spin energy into the external shock.~During the prompt phase,
the newly born magnetar strongly interacts with the accretion disc left behind by the progenitor, 
and may undergo transitions to a propeller regime \citep{illa75, stella86}, where disc material cannot enter the magnetosphere and the accretion power is reduced.~We thus set to study whether this scenario is compatible with the general luminosity evolution observed in prompt GRB lightcurves, which can also explain the $B-P$ correlation found by \cite{stratta18}.~Previous suggestions to interpret GRB precursors in a similar framework gave encouraging results in a few GRBs~\citep{bernardini13, bernardini14}.

As shown by~\citet{stella86} and \citet{campana98}, for transient X-ray  binaries hosting different populations of NSs,
the onset of propeller is characterized by a distinctive knee in the decaying phase of the light curve, from which the transition luminosity can be determined.~By independently measuring the spin (via their pulsed emission) and magnetic field (via cyclotron absorption lines and/or spindown timing) of central objects in a variety of accreting sources,~\cite{campana18} demonstrated the existence of a universal relation between luminosity, magnetic field and spin period for the onset of propeller, as theoretically predicted (eq.~\ref{eq:Bell-relation} below), spanning many orders of magnitude in each of these quantities.~Their sample included low and high-mass X-ray binaries, cataclysmic variables, and young stellar objects. 

Here we select a sample of GRBs that are promising candidates for having a magnetar engine, in that their afterglow light curves show clear evidence for a plateau.~Among these, we select only the ones with an accurate estimate of the jet opening angle, and with a prompt emission light curve which is well sampled in time, and has a good spectral coverage to allow an accurate determination of the true bolometric luminosity after correcting for beaming.~By measuring the B-field and the spin period from the plateau in the X-ray afterglow, and the transition luminosity from the prompt radiation, we test whether these GRBs verify the same relation as the known accreting sources in propeller.~Through this relation we are able to verify the proposed scenario in which accretion energy powers the prompt emission, and the NS spin energy powers the afterglow plateau once accretion subsides.~Moreover, we are able to 
independently constrain the radiative efficiency of accretion in GRBs. 

Our paper is organized as follows: Sec.~2 describes the basics of the theory and the accretor-propeller transition. Sec.~3 describes the GRB sample, clearly highlighting our selection criteria, and then individually describing the properties of each source.~We connect theory to observations in Sec.~4, where we 
assess the significance of the correlation that we find.~We summarize and conclude in Sec.5.

\section{The accretor-propeller transition}
\subsection{General theory} 
Matter forming an accretion disc flows towards the accretor by transferring angular momentum outwards, and by releasing (half of) its gravitational potential energy.~If the accretor is a magnetic~NS (with mass $M$ and radius $R$), the disc~may not reach its surface, being truncated at the magnetospheric radius, $r_m >~R$, inside which the NS magnetic field controls the dynamics of the inflowing plasma.~In disc geometry $r_m$~is a fraction $\xi$~of~the Alfv\`{e}n radius, $r_A$, at which magnetic pressure balances the ram pressure in a spherical flow (e.g. \citealt{pringle72}, \citealt{frank03})
\be
\label{eq:magnetosph}
r_m = \xi r_A = \xi \left(\displaystyle \frac{\mu^4}{2 G M \dot{m}^2} \right)^{1/7} \, .
\ee
Here $\dot{m}$ is the mass accretion rate, $\mu = BR^3/2$ the NS magnetic moment, $B$~the dipole magnetic field at the pole, and the parameter $\xi \sim 0.5$ (e.g.~\citealt{ghosh79, campana18}) encodes the microphysics of the interaction between the accreting plasma and the $B$-field co-rotating with the NS.

When the disc is truncated at $r_m > R$, two cases can occur: (i) if the disc (Keplerian) rotation at $r_m$ is faster than the~NS rotation, the NS is in the {\it accretor} phase, matter enters the magnetosphere transferring its excess angular momentum to the NS, gradually approaching corotation; (ii)~if the disc rotation is slower, the NS enters the {\it propeller} regime, in which the plasma experiences a centrifugal barrier at $r_m$ and is either flung out \citep{illa75}, absorbing in the process some of the stellar angular momentum, or piled up near $r_m$ (e.g. \citealt{dangelo10}).

 What differentiates these two cases is the ordering of~$r_m$ and of the corotation radius, $r_{\rm co}$, defined as the distance from the NS at which the disc rotation equals the NS angular frequency ($\omega$), $r_{\rm co}=(GM/\omega^2)^{1/3}$.~In the accretor regime, $r_m < r_{\rm co}$, matter falls all the way to the NS surface releasing its remaining gravitational potential energy and producing the accretion luminosity $L = \epsilon GM \dot{m}/R$, where we accounted for a radiative efficiency  ($\epsilon \le 1$).~{The corresponding, frequently used conversion efficiency of rest mass,} $\eta$ in the relation $L = \eta \dot{m} c^2$,~is $\eta = \epsilon GM/(c^2R)$.~In the propeller regime, $r_m > r_{\rm co}$, accretion is halted at $r_m$ and the source has a lower luminosity, 
$L_{\rm prop} = \epsilon \dot{m}(GM/2 r_m)$, at the same mass inflow rate.  

The minimum accretion luminosity, $L_{\rm min}$, is obtained by first equating $r_m$ to $r_{\rm co}$: the resulting mass accretion rate, $\dot{m}_{\rm min}$, is fed into the expression for $L$, yielding
\begin{eqnarray}
\label{eq:Bell-relation}
\log_{10} L_{\rm min} + \frac{7}{3}\log_{10}P+ \log_{10}R_6 ~= 37.15 + 2 \log_{10} \mu_{30}~+& &~~~~~ \nonumber \\ 
+\log_{10} \epsilon + \displaystyle \frac{7}{2} \log_{10} \left(\displaystyle\frac{\xi}{0.5}\right) - \frac{2}{3} \log_{10} M_{1.4} ~~~~~~~~~~~~~~~~~~~~~~~~~&~ & 
\end{eqnarray}
where $Q_x =Q/10^x$ in c.g.s. units, $M_{1.4} = M/(1.4 M_\odot)$, and $P= 2\pi/\omega$ is the NS spin period (in seconds).

Conversely, the maximum propeller luminosity is obtained by plugging $r_{\rm co}$ and $\dot{m}_{\rm min}$ in the expression of $L_{\rm prop}$, 
\be
L^{\rm (max)}_{\rm prop} = \frac{\epsilon \dot{m}_{\rm min}}{2} \left(GM \omega\right)^{2/3}= \frac{\sqrt{2}}{4}  \displaystyle \frac{\epsilon \xi^{7/2} \mu^2 \omega^3}{GM}\, .
\ee
The ratio ${L}_{\rm min} / { L}^{\rm (max)}_{\rm prop}  \approx 250 M_{1.4}^{1/3} R^{-1}_6 P^{2/3}$, depends mainly on the NS spin (\citealt{corbet96}).~The corresponding luminosity drop, 
which is typically used as a signature of the propeller transition, is expected to be small for millisecond spins and smoothed out by the finite time of the transition, thus hardly recognizable in the rapidly variable GRB prompt emission.~Theory predicts another transition, the onset of the so-called "ejector", when the disc edge extends farther out, approaching the light cylinder at $r_L = c / \omega$;~then the source luminosity becomes dominated by the NS spindown, which for ideal magnetic dipole radiation is (\citealt{spitkovsky06})
\be
\label{eq:spindown}
L_{\rm sd} = \frac{\mu^2}{c^3} \omega^4 (1+\sin^2 \alpha) \, ,
\ee
\begin{figure}[ht]
\centerline{\includegraphics[scale=0.43]{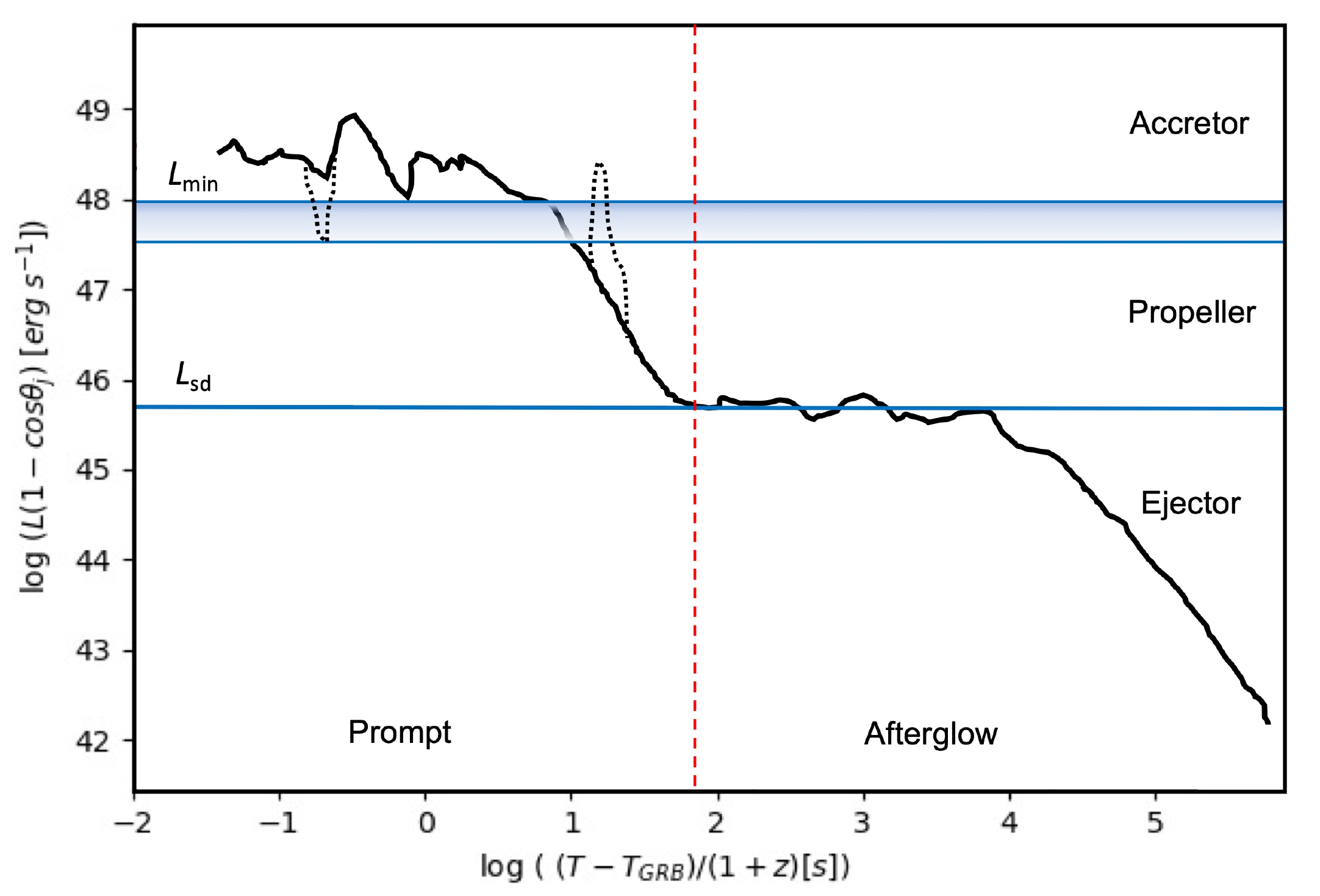}}
\caption{Sketch of 
a GRB lightcurve in which an accreting magnetic NS switches from accretion to propeller while in the prompt phase, and then to the ejector regime when the afterglow~sets~in. 
The vertical span of the shaded band that separates the accretor~from the propeller
phase corresponds to the gap between the minimum accretion luminosity, $L_{\rm min}$, and the maximum propeller luminosity, $L_{\rm prop}^{\rm (max)}$.~This gap is expected to be small ($\sim$ a few) for millisecond spins.~Possible 
short-term transitions that may take place from the accretion to the  propeller regime or vice versa are indicated with a dotted dip or peak in the light curve.~The onset of the ejector regime is marked by the early NS spindown luminosity, $L_{\rm sd}$.~Values on the X and Y axes are roughly indicative of those of (long) GRBs. 
}
\label{fig:skema}
\end{figure}
$\alpha$ being the angle between the magnetic and rotation axes of the NS.~Combining eqs. \ref{eq:Bell-relation} and \ref{eq:spindown} for $\alpha=\pi/2$ gives the ratio 
\be
\label{eq:ratio}
\kappa = { L}_{\rm min}/{ L}_{\rm sd} \approx 1.2 \times 10^5 \epsilon P^{5/3}\left(R_6 M_{1.4}^{2/3}\right)^{-1} (\xi/0.5)^{7/2} \, .
\ee
This ratio too is a function of the NS spin alone\footnote{Eq.~\ref{eq:ratio} assumes the same NS spin at $L_{\rm min}$ and at the ejector onset.} (for given NS mass and radius) but, in contrast to $L_{\rm min} / L^{\rm (max)}_{\rm prop}$, it is robustly determined in our case, as the NS spindown luminosity is well tracked by the luminosity of the afterglow plateau.
\subsection{Peculiar properties of accretion in GRBs}
\label{sec:peculiar}
Fig.~\ref{fig:skema} depicts the lightcurve shape expected as a result of~the above-mentioned transitions.~The theory has been developed for and verified in accreting~NSs in High-Mass X-ray Binaries (HMXBs) Be transients (\citealt{stella86}), Low-Mass X-ray Binaries (LMXBs; \citealt{stella94, campana98, campana98b}), cataclismic variables (CVs) and young stellar objects (YSOs; \citealt{campana18}), which established the universal character of eq.~\ref{eq:Bell-relation}.~Compared to ~such systems, disk accretion in GRBs is expected to present important differences. 

In all the sources studied by \cite{campana18}, the disk material that enters the magnetosphere of the accretor is channeled onto its surface where it releases the gravitational potential energy from $r_m$ to $R$.~The emission we receive comes from the accretion column, close to the stellar surface, with a typical conversion efficiency of energy to radiation~$\epsilon \sim~1$. 

In our proposed scenario the GRB prompt emission is still~a proxy of the mass accretion rate, yet it comes from shocks in a relativistic jet, much farther away from the central engine. 
Thus, GRB lightcurves do not show the characteristic pulsations of accreting NSs signalling their spin periods\footnote{See, however, \citet{chirenti23}.}, and their (featureless) spectra do not offer means for directly measuring the NS $B$-field, in contrast to what occurs in other magnetic accreting stellar objects (e.g., \citealt{campana18}).~Furthermore, radiation is believed to come from the conversion of the jet's kinetic energy with an intrinsically low efficiency ($\epsilon_\gamma \lesssim 0.1$)~and the gravitational potential energy released by accretion is only partially converted into the jet kinetic energy ($\epsilon_{\rm jet} < 1$).~Thus, it is generally expected that $\epsilon = \epsilon_{\rm jet} \epsilon_\gamma \ll 1$.

We will not deal with the details of the interaction between the disk plasma and the ms-spinning NS magnetosphere or with the mechanism by which a relativistic jet is launched.~We~simply assume that the jet will be launched, powered by the energy of accretion and with the known properties of GRB jets,  both during the accretion and propeller phases.

While the relativistic jet carries only a minor amount of baryons, the characteristic mass accretion rate envisioned in the collapsar scenario is $\sim (0.001-1) M_\odot$s$^{-1}$, implying a total accreting mass $\Delta M \sim (0.001-10) M_\odot$ for typical prompt durations $\sim 1-30$ s.~Most of the latter  
may be entrained~in~a slow-moving and wide outflow or end up onto the NS surface, thus releasing its potential gravitational energy and increasing the NS mass.~In this latter case, at the typical GRB mass accretion rates, the infalling material will form a very hot and high-density accretion column in which neutrino emission is dominant 
(e.g.~\citealt{piro11}, \citealt{metzger18}) while the escaping high-energy radiation is negligible.

Finally, because the maximum gravitational mass allowed by the NS EOS is $M_{\rm g, max} \gtrsim 2.2$~M$_\odot$~\citep{cromartie20}, the mass increase of the NS may impact its stability against gravitational collapse.~In case of core-collapse, if the mass~of the NS remnant is $\sim 1.4$ M$_\odot$, like in the observed NS population \citep{kiziltan13}, then there is significant room for accretion before collapse sets in.~In short~GRBs, on the other hand, 
the merger remnant is likely to be close to, or above, the maximum NS mass.~The probability of a stable NS would depend critically on both the initial component masses in the binary progenitor and the NS EOS (e.g.~\citealt{dallo15,piro17,margalit17}). 
\subsection{Main observables}
\label{sec:observables}
If the GRB central engine is a millisecond-spinning magnetar, we expect to observe in the prompt light curve the characteristic steep flux decays of switching-off relativistic jets (e.g.~\citealt{kumar15}) when the NS enters the propeller regime.~
The minimum accretion flux $F_{\rm min}$, and hence $L_{\rm min}$, can thus be estimated directly from the data (as described in Sec.~\ref{sec:data}).~Within the magnetar scenario it then becomes possible to verify, in a way that parallels the method of \cite{campana18}, whether the putative NSs in GRBs follow the universal relation (\ref{eq:Bell-relation}). 

\begin{figure*}[t]
\centering
\includegraphics[scale=0.365]{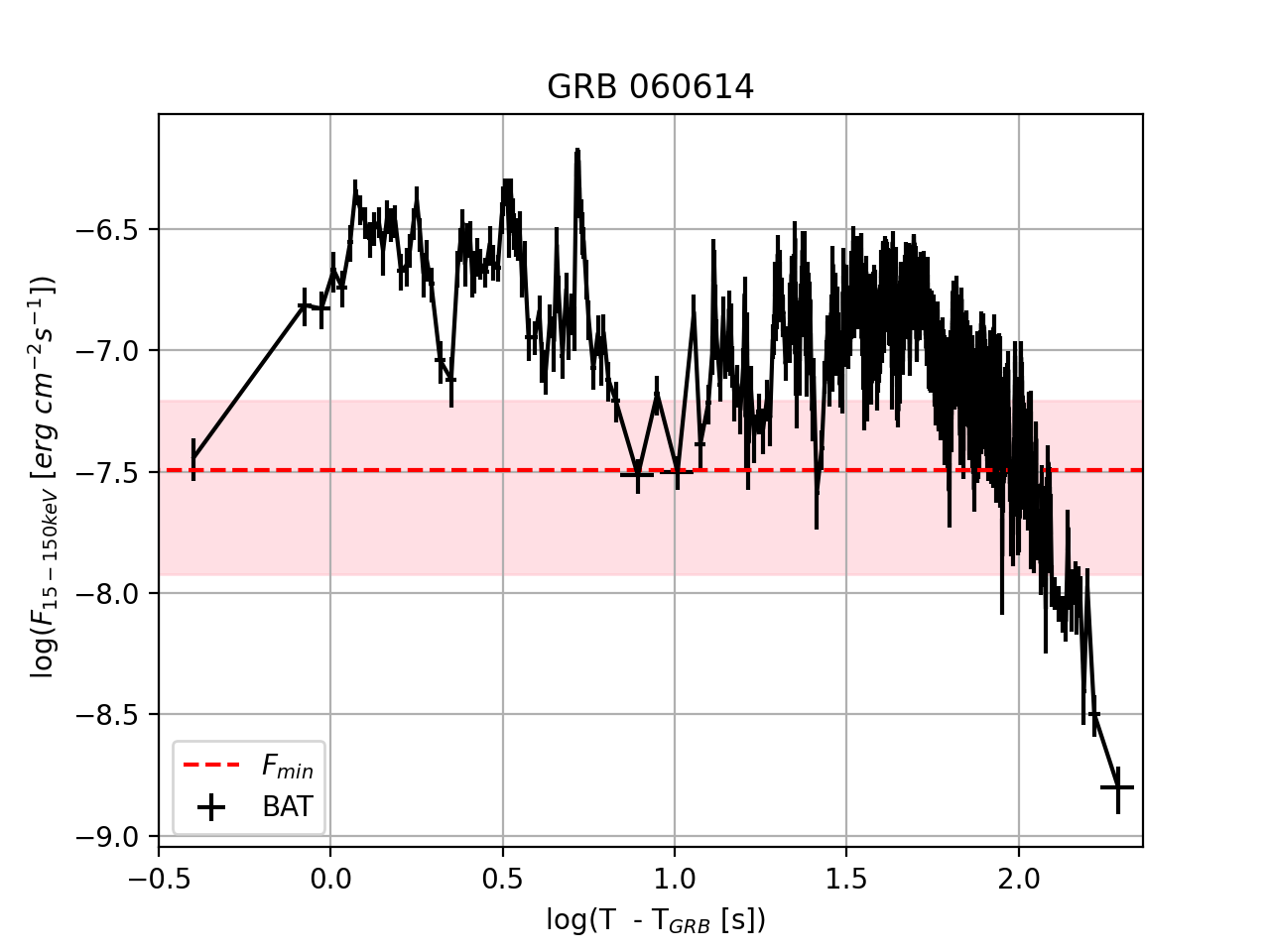}
\includegraphics[scale=0.365]{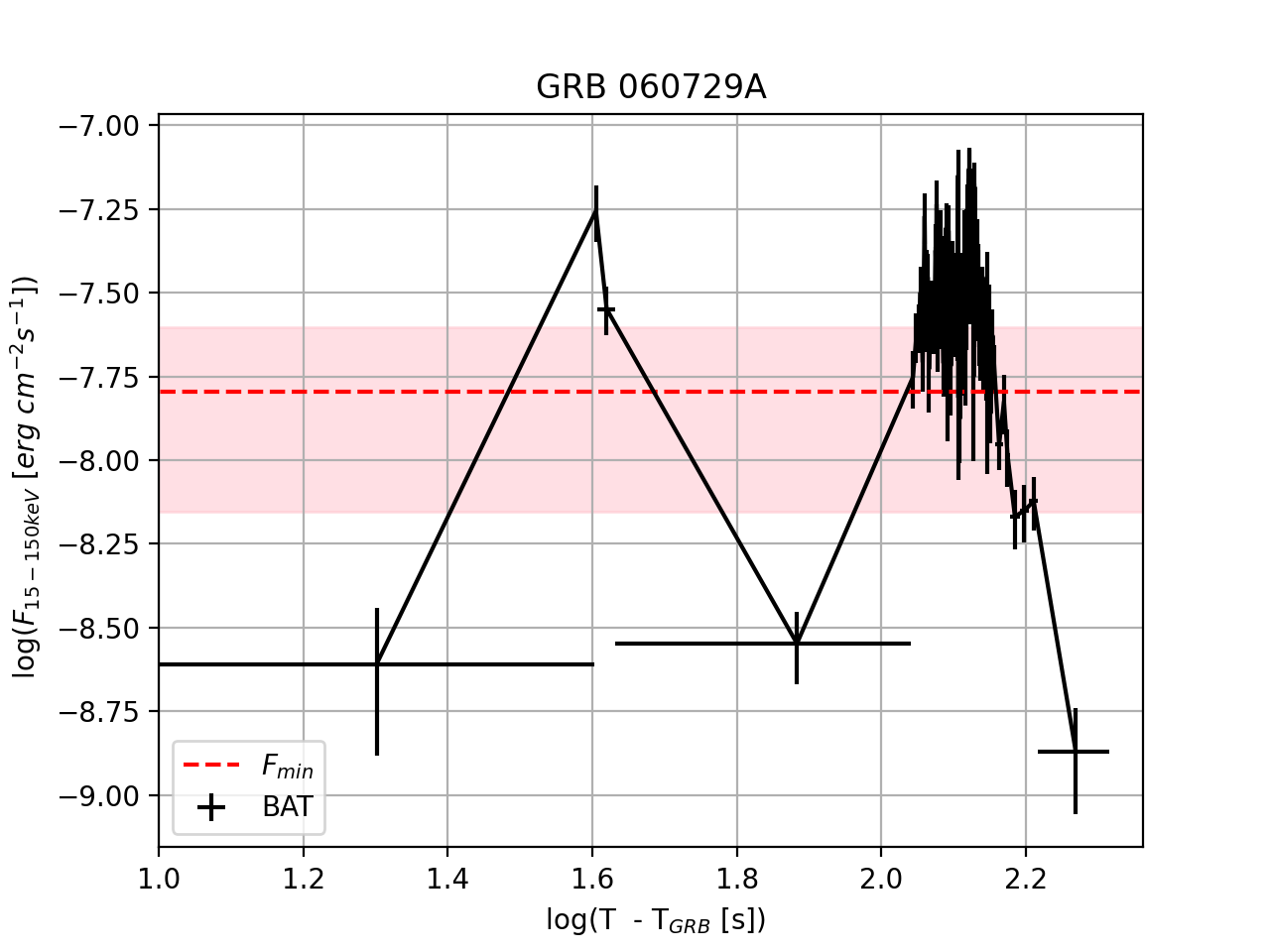}
\includegraphics[scale=0.365]{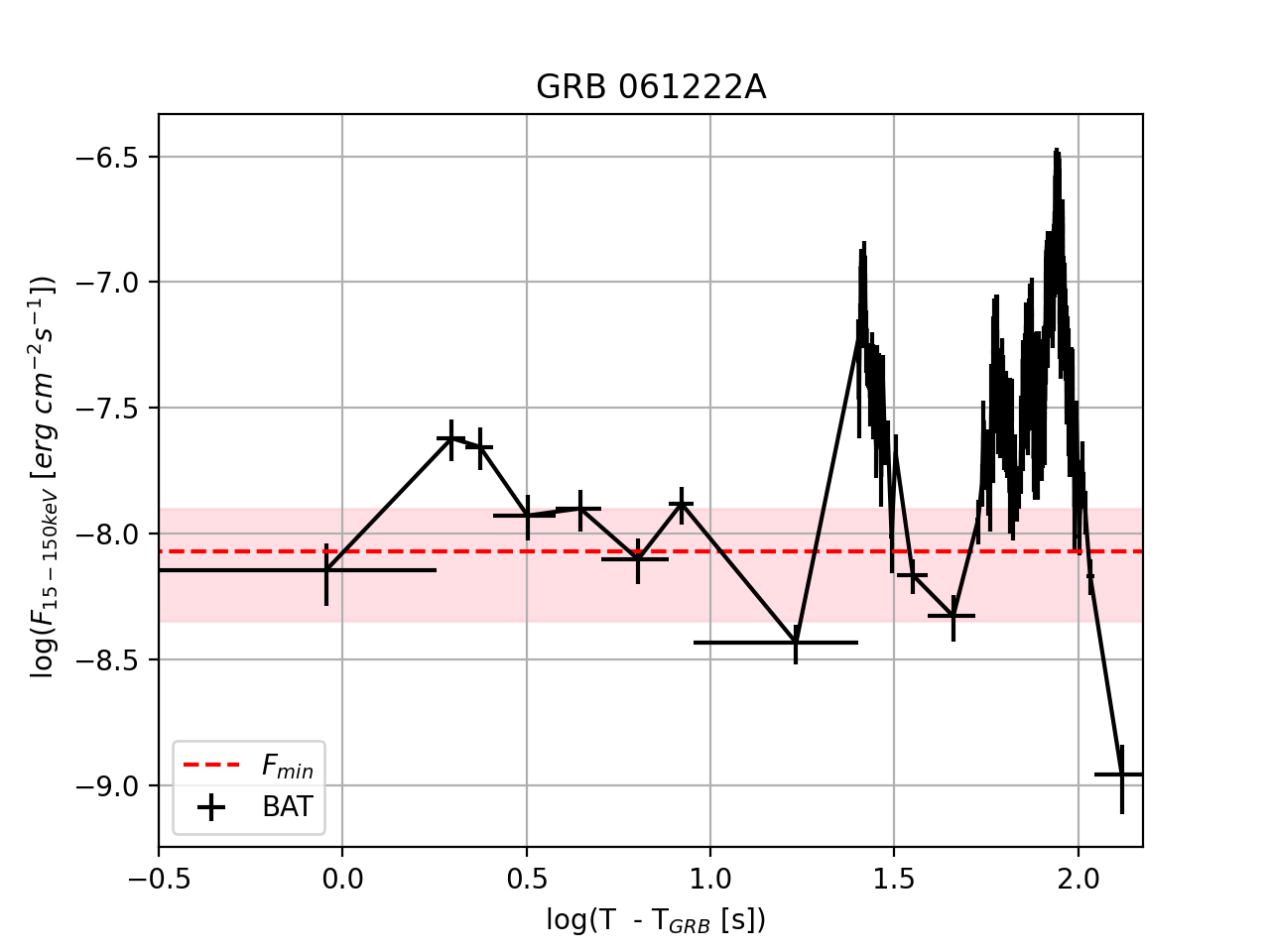}
\includegraphics[scale=0.365]{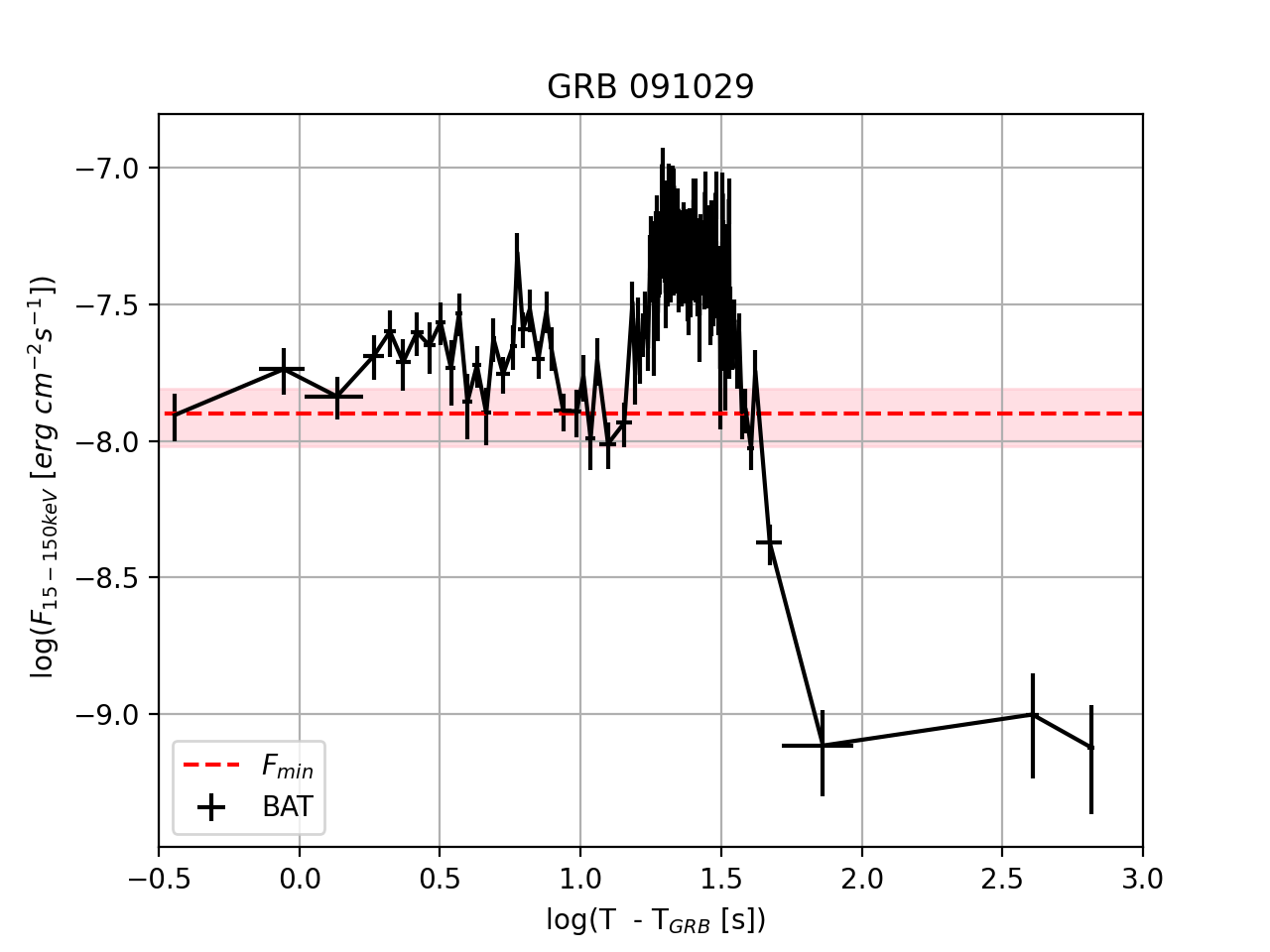}
\includegraphics[scale=0.365]{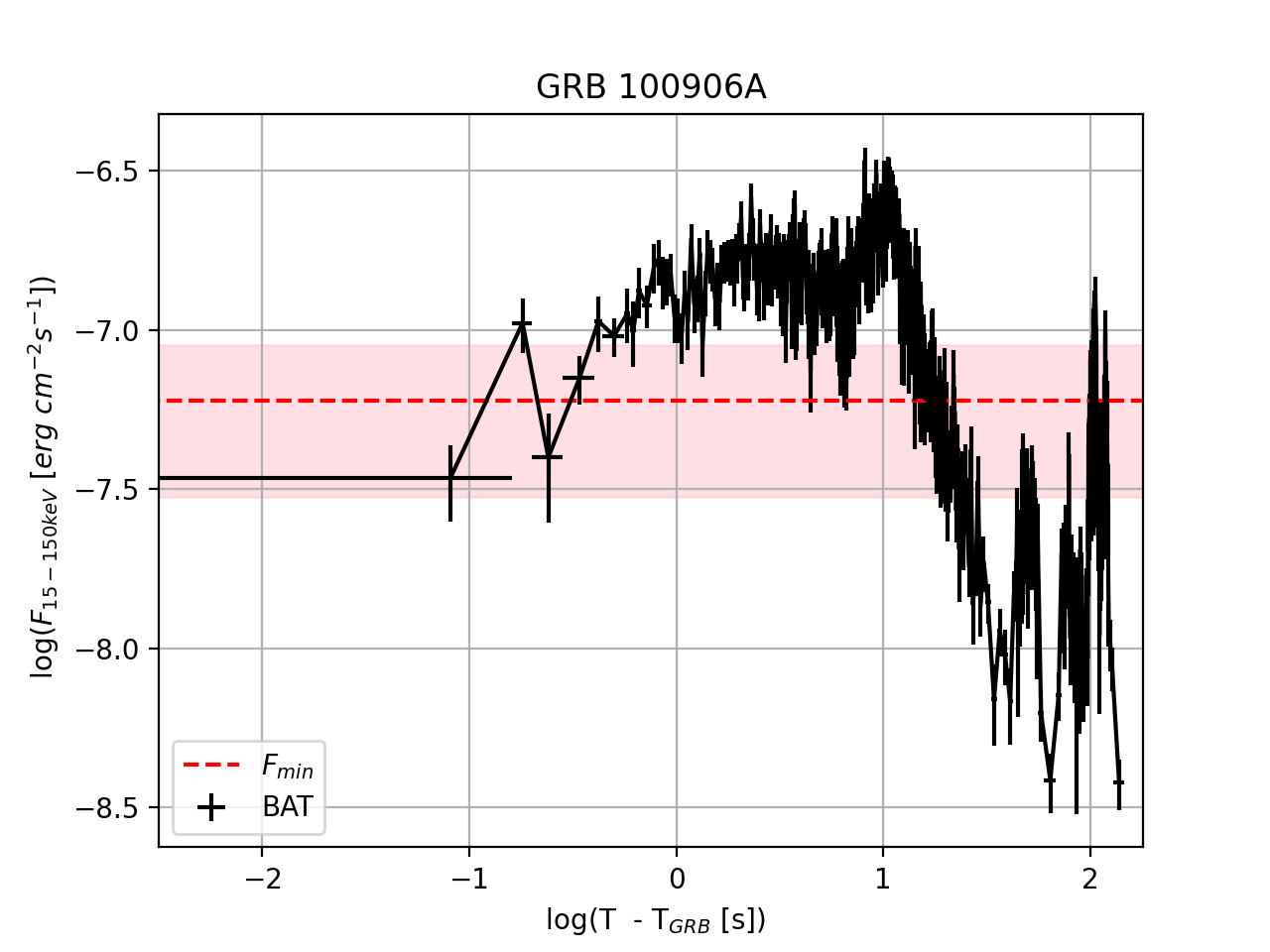}
\includegraphics[scale=0.365]{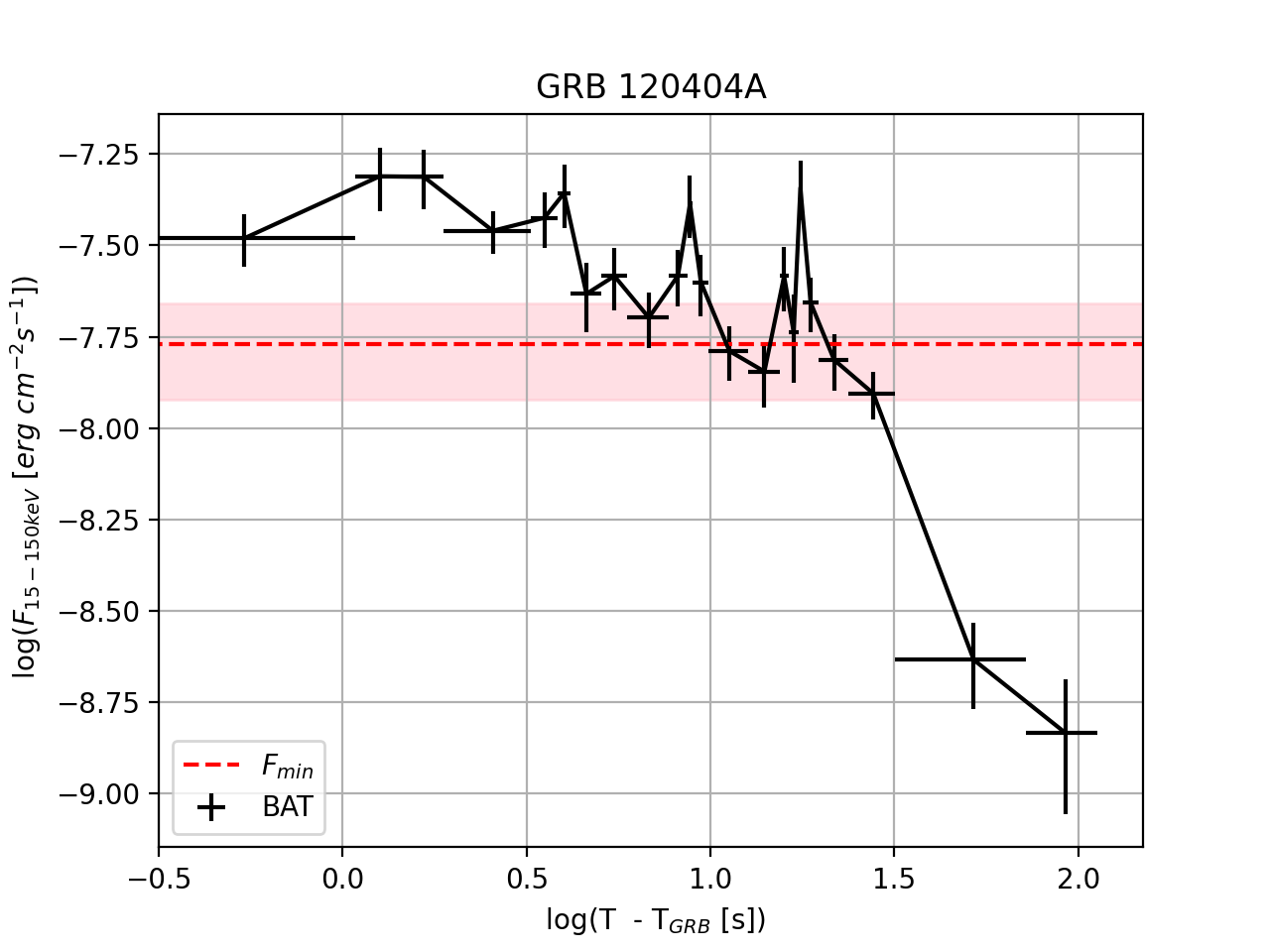}
\includegraphics[scale=0.365]{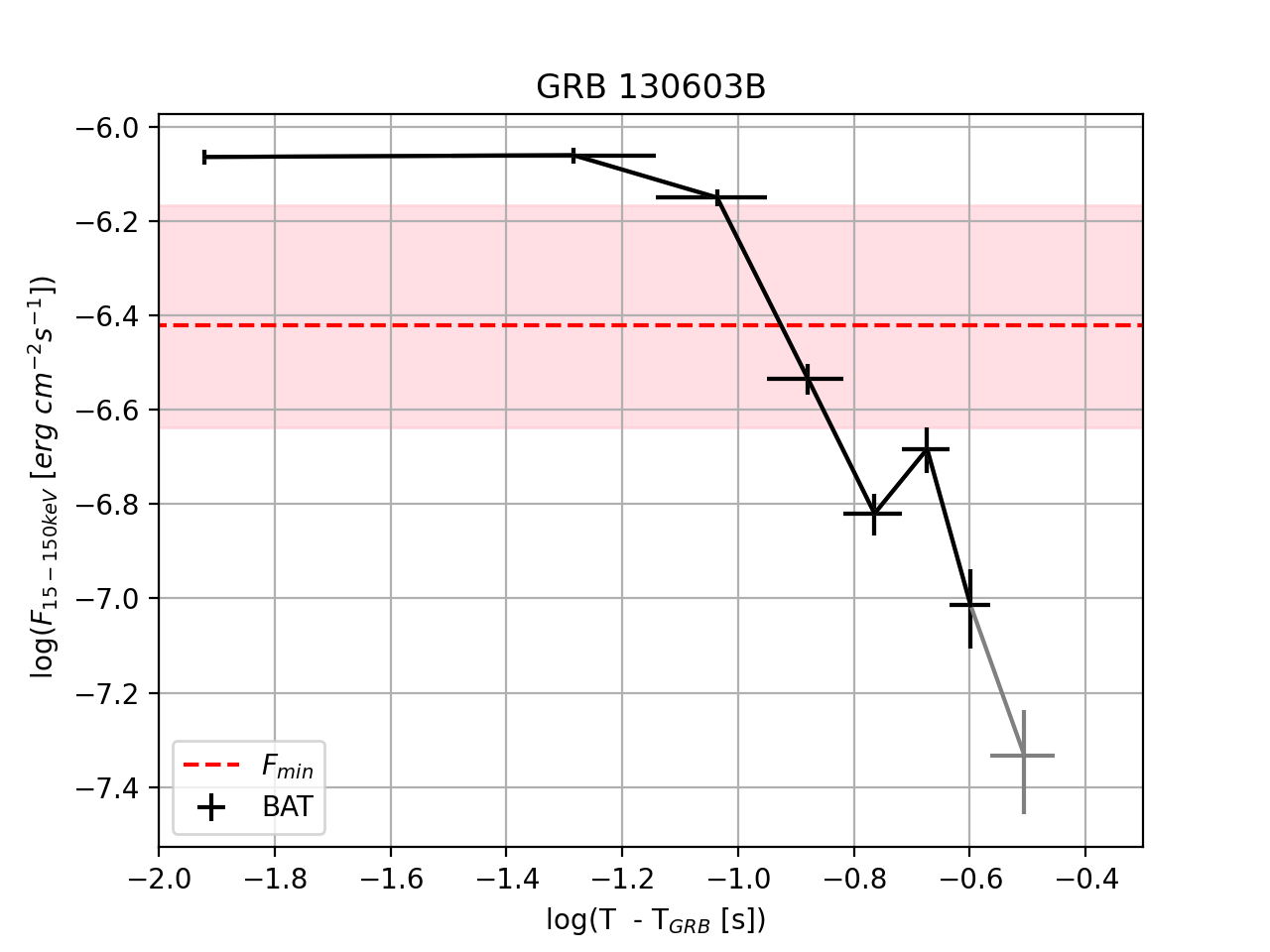}
\includegraphics[scale=0.365]{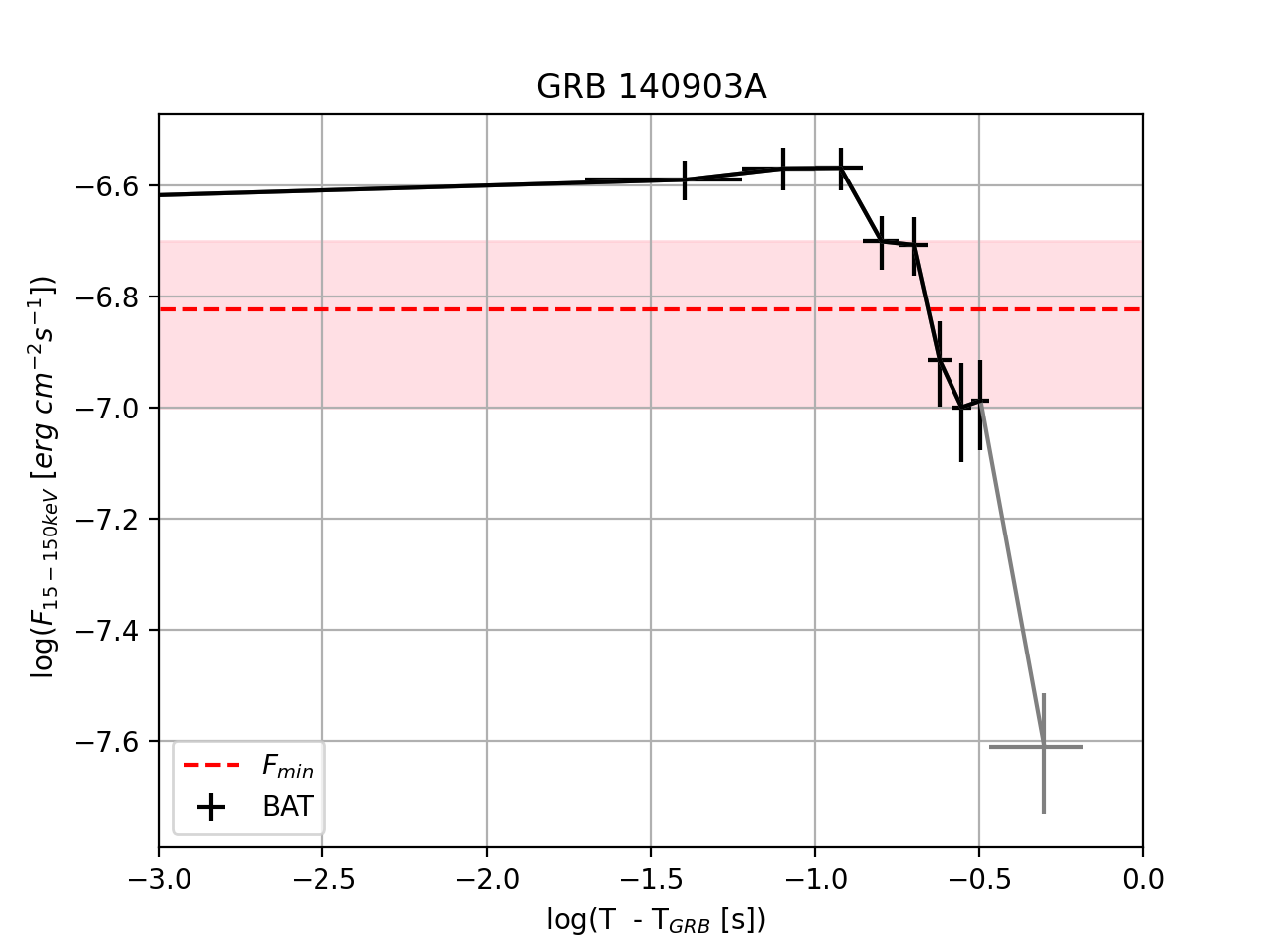}
\includegraphics[scale=0.365]{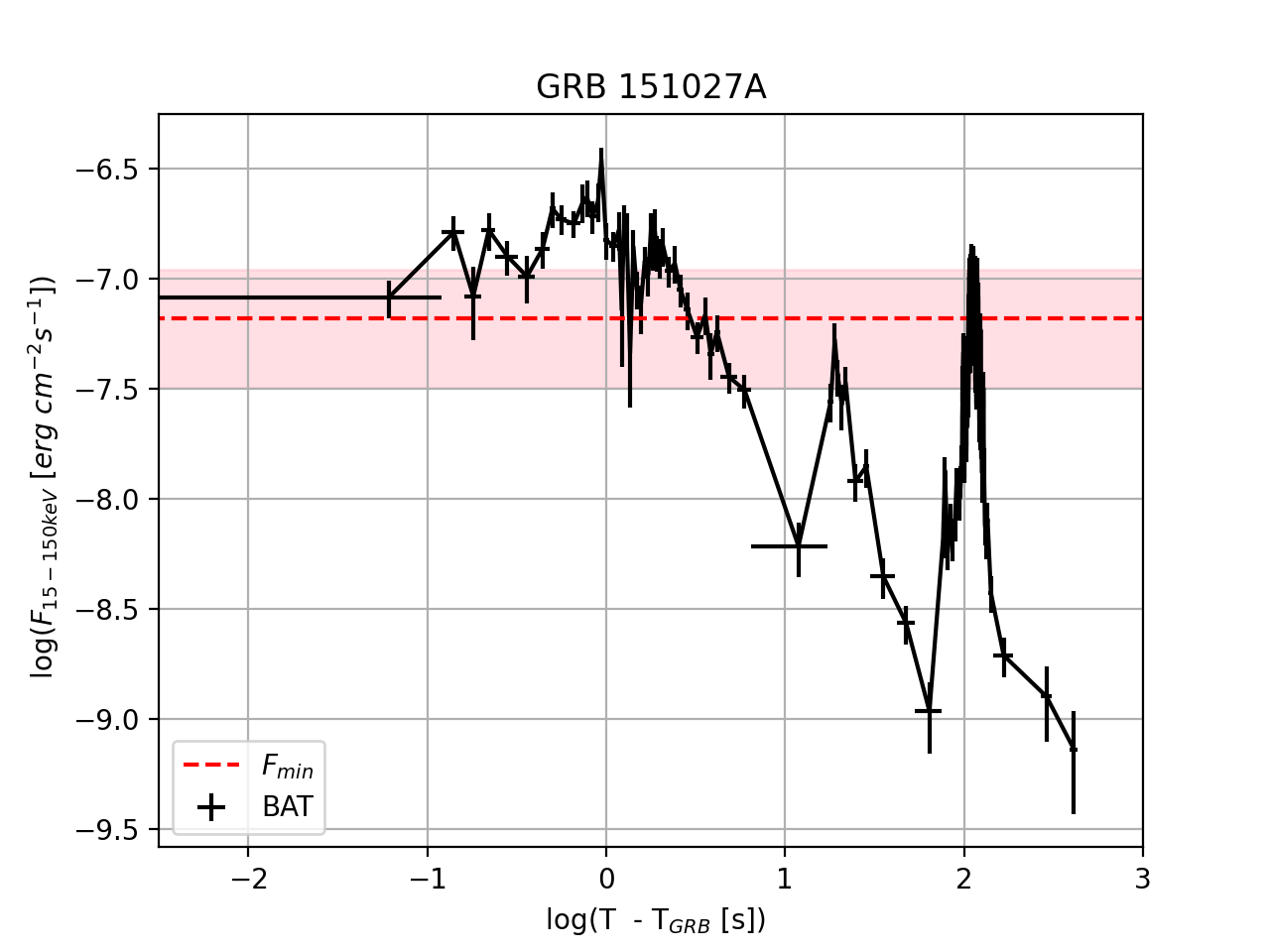}
\caption{
Prompt gamma-ray observed flux light curves from Swift/BAT (15-150 keV) data (with signal to noise ratio of 5) for  the GRBs in our sample.~  
For each burst, the chosen value of $F_{\rm min}$ is indicated by the red, dashed horizontal line while the pink shaded area indicates the uncertainty on $F_{\rm min}$, as described in $\S$~3.1. The grey data points at the end of the two short GRB prompt light curves indicate the flux level obtained from data with signal to noise ratio of 4.}
\label{fig:batlc}
\end{figure*}

With this goal in mind, we searched the prompt light curves of the GRBs in our sample for prolonged and steep flux decays, spanning at least one order of magnitude, in~analogy to what is typically done in other classes of accreting~compact objects (e.g. \citealt{campana18} and references therein).~Assuming these drops are associated to the propeller onset, we set $F_{\rm min}$ to be 
limited by the lowest minimum and the lowest maximum prior to the flux decay or by the occurrence, after the decay, of a new peak signalling a new temporary transition to~the accretor phase.~With these criteria we defined an allowed~range for $F_{\rm min}$,~and set the fiducial values at the midpoint of the~range.~In $\S$ \ref{sec:data} we provide more details on this procedure for each GRB.~Our results, illustrated in Fig.~\ref{fig:batlc}, are reported in Tab.~\ref{tab:tab1}.~Moreover we estimated the NS spin period $P$ at the end of the prompt phase, and its magnetic dipole field strength $B$ by fitting the observed plateau in the X-ray afterglow with a model of energy injection by a spinning down magnetar, as further discussed in sec.~\ref{sec:magnetarspindown} (cf.~\citealt{dallosso11}, \citealt{bernardini13}, \citealt{stratta18}).

Note that 
here we estimate $L_{\rm min}$ directly from the prompt light curve, independently of the $B$- and $P$-values derived from the afterglow, whereas \citet{bernardini13} used the latter values to infer $L_{\rm min}$ in the prompt phase.~Ultimately, this allows us to verify~the validity of eq.~\ref{eq:Bell-relation} using independent quantities, while at the same time constraining the radiative efficiency, $\epsilon = \epsilon_{\rm jet} \epsilon_\gamma$. 

\section{Data analysis}
\label{sec:data}
We selected a sample of both long and short GRBs detected with the Swift satellite and at known distances that present (i)~clear evidence of an X-ray plateau in the afterglow light curve,~(ii) a well sampled prompt emission light curve,  to allow the study of flux variations consistent with transitions to the propeller phase, 
(iii)~a good spectral coverage of the prompt emission with a measure of the peak energy, (iv) an accurate estimate of the jet opening angle $\theta_j$ from multi-band afterglow data.~Meeting these four criteria is challenging.~For this reason, the resulting sample is made of only 9 events, 
wo of which are short GRBs (140903A, 130603B; see Tab.~1). 
\subsection{Estimating $F_{\rm min}$}
To estimate the minimum flux level $F_{\rm min}$ associated with the accretion phase, we used the publicly available Swift/BAT 15-150 keV prompt light curves\footnote{From the Swift/XRT Burst Analyzer repository (\citealt{Evans10}).} with temporal resolution that guarantees a signal to noise ratio $>5$.~Following \S2.3, for each 
prompt light curve, we identified a maximum and minimum flux within which we could confidently constrain the propeller transition.~Specifically, the maximum flux was associated with the faintest of the peaks that characterise the bright prompt phase, while the minimum flux with minima prior to the steep flux decrease.~Results are quoted in Table~2.

Below we describe, for each GRB, the prompt emission features based on which we estimate $F_{\rm min}$, as well as the studies from which the $\theta_j$-values were adopted.
\subsubsection{GRB~{\rm 060614}}
This  peculiar GRB at $z=0.125$ has a 
duration, $T_{90}=102$~s, suggesting a long GRB classification, while the lack~of an associated supernova down to very deep limits, and its location in the time-lag/peak-luminosity plane, are more consistent with the properties of short GRBs \citep{mangano07}.

The prompt emission light curve is very well sampled,~the X-ray afterglow has a clear plateau and an excellent multi-band data set is also available.~The jet opening angle was taken from \citet{mangano07} that quote $\sim10.5$ deg from a very robust achromatic temporal break observed in the afterglow light curve 117 ks after the burst onset. 

The prompt light curve shows two bright intervals separated by a hollow minimum at $\lesssim 10$~s.~Starting at $\sim 50$~s post-trigger, an extended and steep flux decay is observed.~Prior to it, the hollow minimum and the lowest peak at $\lesssim 10$~s indicate a narrow range of possible $F_{\rm min}$-values, consistent with the flux in the first part of the steep decay.~We extended this range to  include even the lowest peak at $\gtrsim 100$~s, a choice that has only a minor effect on the fiducial value of $F_{\rm min}$.
\subsubsection{GRB~$060729A$}
The Swift/BAT light-curve of this GRB, at z=0.543, shows clear evidence of a precursor followed by a long quiescence time ($\sim$100s), which is ended by a typical burst lasting $\sim40$s. 
The X-ray light curve of its afterglow is characterized by~a long plateau\footnote{Incidentally, a NASA press release of this burst highlighted its peculiar plateau, suggesting a magnetar central engine \url{https://www.nasa.gov/centers/goddard/news/topstory/2007/gammaburst_challenge.html}}  ($\gtrsim 4 \times 10^4$ s) and post-plateau power-law decay, with a very late break  \citep{racusin09}.~From the X-ray lightcurve alone (no multi-band data available) these authors classified GRB 060729A as a ``prominent jet break" case, and measured its $\theta_j$ accounting for the effect of prolonged energy injection, which may extend the power-law decay beyond the time expected in the standard scenario.~Given the accurate determination of $\theta_j = 6.5^{+0.25}_{-0.15}$~deg, we included this event in our sample despite the lack of multi-band data. 

The $F_{\rm min}$-range in this GRB is constrained between the lowest peak at $\approx$ 160~s and the flux at the onset of the steep flux decay; the latter feature appears to be repeated twice in the Swift/BAT light curve (upper central panel in Fig.~\ref{fig:batlc}).

\subsubsection{GRB 061222A}
This is the brightest long GRB in our sample.~Its ~light curve was recognized by \cite{bernardini13} as a prototypical example of the way in which the propeller mechanism can switch a GRB central engine on and off.~The prompt phase, lasting  $ \sim 150$ s, is characterized by three main pulses of which the latest is the brightest. 
 We identified $F_{\rm min}$ as being contrained between the lowest peaks and the lowest minima observed prior to the steep flux decays occurring at $\lesssim$ 30~s and at $\lesssim$ 100~s.

An half-opening angle $\sim~2.4$~deg was inferred, like for GRB 100906A, in the multi band afterglow analysis by \citet{chandra12}.~The more recent and detailed study by \citet{ryan15} quoted a similar result, providing  $\theta_j=  3.7^{+0.8}_{-0.4}$ deg, the value adopted here.

\subsubsection{GRB$~091029$}
This bright long burst has a prompt duration $T_{90}=(39\pm5$)~s in the Swift/BAT range \citep{barthelmy09}.~At $\sim$ 35 s we identify the characteristic flux drop 
indicating the onset of the propeller regime.~Comparing with the lowest peaks and minima prior to this transition (in particular at $T-T_{\rm obs} < 10$~s), we determined a narrow range for $F_{\rm min}$ with a central value of $F_{\rm min}\approx 1.3\times10^{-8}$~erg~cm$^{-2}$~s$^{-1}$.

The afterglow of this GRB was monitored in several bands (e.g. \citealt{filgas12}).~The optical light-curve shows~a few rebrightenings that challenge the measure of the jet break epoch in this band. 
\citet{lu12} determined $\theta_j\sim0.058$ rad ($\sim3.3$ deg) in a Swift GRB sample study on the selection effects in the apparent $\theta_j-z$ dependence. \citet{ryan15} obtained $\theta_j = 12^{+11}_{-8}$ deg, and a viewing angle $\theta_{\rm v} = 0.75^{+0.18}_{-0.47} \times  \theta_j$, modelling the X-ray afterglow by means of a high-resolution hydrodynamical model with no angular structure in the jet (top-hat), and imposing a narrow prior on $\theta_j \sim \left[2.6-28.6\right]$ deg.

By studying the optical/X-ray afterglows of a sample of GRBs, \citet{wang18} identified an achromatic temporal break at ($30.0\pm 6.8$) ks for this burst.~By assuming a constant density circumburst medium (ISM), and the standard jet-break formula  (e.g. eq.~1 in \citealt{sari99}),
an half-opening angle  $\theta_j=1.24\pm0.92$~deg was inferred. 
In Fig.~\ref{fig:universe_mag90} and in the next sections ~we adopt this result, which was obtained from the identification of an achromatic break in a multi-band study.

\begin{table*}[t]
\footnotesize
\centering
\begin{tabular}{|c|c|c|c|c|c|c|c|c|}
\hline
GRB & $z$ & $T_{90,100}$ & $F(E)$ &  $E_{\rm peak,obs}$ & $\alpha$ & $\beta$ & $K$ & $E_{{\rm iso}, 51}$ \\
 ~   &   &    [s]  &  & [keV]    & &    & & 1/(1+z)~keV-10MeV\\
\hline
091029 & 2.75 
& 50 & BAND & 66 & -1.56 & -2.38 & 2.39 & 156.5  \\
100906A & 1.73 
& 90 & CPL & 195 & -1.6 & -- & 2.54 & 249 \\
151027A & 0.81 & 117 & CPL & 173 & -1.44 & -- & 2.23 & 33.0 \\
061222A & 2.09 
& 60 & CPL & 298 & -0.89 & -- & 2.95 & 259.9 \\
120404A & 2.88 
& 43 & CPL & 70 & -1.61 & -- & 2.09 & 103.1  \\
060729A & 0.54 
& 126 & BAND & 8.7 & -1.14 & -2.05 & 2.94 & 58  \\
060614 & 0.13 & 123.6 & CPL & 76 & -1.92 & -- & 2.82 & 2.7  \\
\hline
130603B & 0.36 
& 0.07 & CPL & 607 & -0.67 & -- & 6.64 &  1.96  \\
140903A$^a$ & 0.35 
& 0.3 & BAND & 40 & -1 & -2 & 3.14 & 0.06 \\
\hline
\end{tabular}
\caption{Prompt emission properties taken from the Swift/Konus-Wind GRB catalog (Tsvetova et al.  2017, 2020) for the GRBs of our sample.~The best fit prompt emission spectral model $F(E)$ is taken as the one corresponding to the integrated spectra (flag 'i'), where BAND is the Band model \citep{band93} and CPL is the cut-off power law model with cut-off energy $E_{\rm cut} = E_{\rm peak}/(2+\alpha)$ and $\alpha$ is the power law photon index. The last two lines indicate the two short GRBs present in our sample. \\ 
$^a$ GRB 140903A is the only one not present in the Swift/K-W catalog and we assumed a BAND model with fiducial values (see text).}
\label{tab:tab0}
\end{table*}

\subsubsection{GRB 100906A}
This is a long GRB among the brightest of our sample.~Its Swift/BAT prompt duration is $T_{90}=(114.4\pm1.6)$~s \citep{barthelmy10} and its light-curve shows a main bursting episode followed by a flux decay and then by multiple lower peaks (Fig.\ref{fig:batlc}).~The steep flux decay at $t \approx (10-30)$~s marks the onset of the propeller regime, from which we estimate a fiducial value of 
$F_{\rm min}\sim 6 \times 10^{-8}$ erg cm$^{-2}$s$^{-1}$, further supported by the later peaks, at $t\approx 120$ s, reaching a similar flux.

The jet half-opening angle $\theta_j=2.9$ deg for this burst is taken from the multi-band afterglow study by \citet{chandra12}, who assumed a constant density circumburst environment following \citet{sari99}.

\subsubsection{GRB 120404A}
The prompt duration of this long GRB is $38.7\pm4.1$ s in~the 15-150 keV Swift/BAT band \citep{ukwatta12}.~The~light curve shows a mildly fluctuating flux up to $\sim$ 20~s, when a steep flux decay begins.~Peaks and dips in the early 20~s constrain $F_{\rm min}$ in a range that is perfectly consistent with the turning point of the flux decay at $F_{\rm min} \approx 2\times 10^{-8}$ erg cm$^{-2}$~s$^{-1}$. 

\citet{guidorzi14} measured $\theta_j\sim 23$~deg fitting broadband, high–quality optical/NIR data with the hydrodynamical code of \citet{vaneerten12}.~Their fit, which did not include energy injection, was found to underpredict the X-ray flux compared to the optical, and required a very steep power-law slope $p \sim 3.4$ for the electron energy distribution.

\citet{laskar15} described the broadband afterglow light curve (radio, NIR, optical, X-rays) with a phenomenological energy injection model.~They obtained $\theta_j=3.1 \pm0.3$~deg~and $p\approx 2.1$, interpreting the optical/X-ray decline $\sim t^{-2}$ seen at $\sim$ 0.1 d as indicative of a jet break.~For these reasons we adopt in the following this latter value of $\theta_j$.

\subsubsection{GRB 130603B}
For this short GRB at $z=0.36$ a kilonova component was identified 
in the optical/NIR afterglow $\sim$~one~week after the event \citep{tanvir13}, consistent with a binary NS merger.~ The possible range of $F_{\rm min}$ is limited from above by the flux at the onset of the steep flux decay, at $T-T_{\rm obs} \approx 0.085$~s, and from below by the occurrence of a late peak  at $T-T_{\rm obs} \approx 0.2$~s.~Accordingly, we determined a fiducial value $F_{\rm min} \approx 3.8 \times10^{-7}$ erg cm$^{-2}$~s$^{-1}$. 

The jet half-opening angle was measured from  multi-band jet break observations as $\theta_j \sim 4-8$ deg assuming a constant density circumburst medium with $n<1$ cm$^{-3}$ \citep{fong15}.~Adopting the same method and setting $n=1$ cm$^{-3}$, \citet{wang18} obtained $\theta_j= (4.47\pm 0.68)$ deg.

Recently, the afterglow lightcurve was re-analyzed by \citet{aksulu22} by means of ScaleFit hydrodynamical simulations \citep{ryan15},  obtaining $\log \theta_j =$ $-0.97^{+0.28}_{-0.54}$,~{\it i.e.}  $\theta_j = 6.3^{+1.7}_{-5.1}$ deg.  In this work we assume the latter result.

\subsubsection{GRB 140903A}
\label{sec:140903}
This short GRB is more than 20 times less energetic than the other one in our sample (130603B) and at a similar redshift ($z=0.35$).~Its Swift/BAT light curve is roughly constant, with minor fluctuations, up to about $\lesssim 0.3$~s, after which it decays close to the detection limit.~We interpret the early behaviour as being due to accretion at various luminosity levels, and the later decay as due to the onset of propeller.~Accordingly, we set $F_{\rm min}$ at  $\approx 1.5\times10^{-7}$ erg cm$^{-2}$ s$^{-1}$. 

There are no Konus-Wind data for this burst.~However, its Swift/BAT 15-150~keV spectrum is consistent with a power law $F(E)\propto E^{-\Gamma}$, with photon index $\Gamma=1.99\pm0.08$ \citep{troja16}.~The relation $\log(E_{\rm peak})=3.258-0.829\Gamma$, for $1.3<\Gamma<2.3$ (\citealt{sakamoto09}), can thus be used to derive a rough estimate of the peak energy.~Accordingly, we assumed a Band model \citep{band93} with observed peak energy $\sim40$ keV, and spectral indices  $\alpha=-1$ and $\beta=-2$.

The multi-band afterglow monitoring shows indications for an achromatic break at $t \sim 10^5$~s  \citep{troja16}.~Using the hydrodynamic simulations of van Eerten et al. (2015), the multi-band data fitting by \cite{troja16} determined a jet half-opening angle $\theta_j = 5.0 \pm 0.7$ deg.~A 
consistent value, $4.0^{+5.0}_{-1.6}$~deg, was determined by \citet{aksulu22} using a dfferent set of hydrodynamical simulations of structured jets. Here we adopt the latter, more recent result.

\subsubsection{GRB 151027A}
The duration of this long GRB derived from its Swift/BAT lightcurve is $130\pm6$~s \citep{palmer15}.~The bright peak in the first $\sim3$ s is followed by a steep decay up to $\sim$ 10~s, and then by a minor re-brightening between $\sim15$~s and $\sim25$~s,~where it reaches a peak $\sim$ 30\% of the one at 3 s.~About $\sim 60$~s~after the burst trigger the light curve shows a new substantial flux increase, strongly suggestive of a new temporary onset of accretion.~This interpretation is supported by the flux level reached in this second bright episode, which is comparable to the one at $\sim 3$~s.~Therefore, we adopt the flux measured at $3$ s as $F_{\rm min}$, with a value of  $\approx6.6\times10^{-8}$ erg cm$^{-2}$~s$^{-1}$. 

A comprehensive multi-band {study of this bust was carried out assuming a wind environment, as suggested by late radio data, deriving $ \theta_j \sim6.3$~deg} \citep{nappo17}.

\begin{figure*}[ht]
\centering
\includegraphics[scale=0.365]{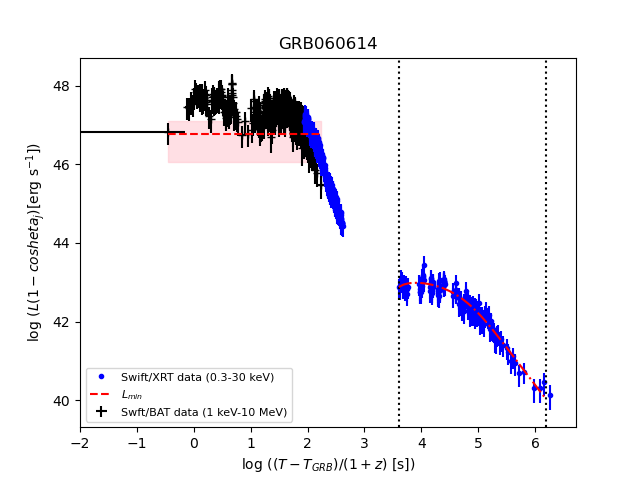}
\includegraphics[scale=0.365]{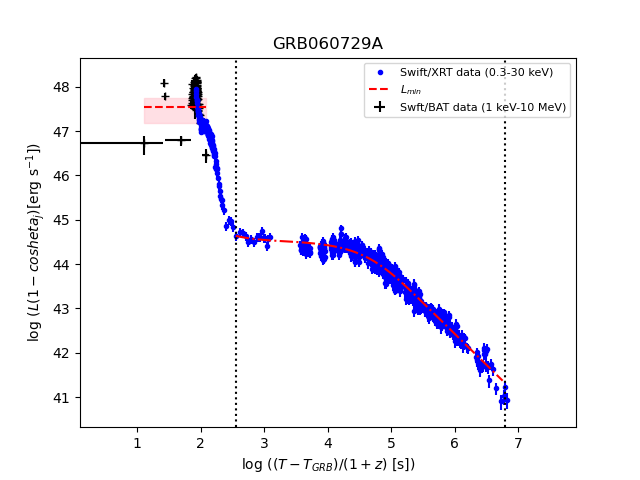}
\includegraphics[scale=0.365]{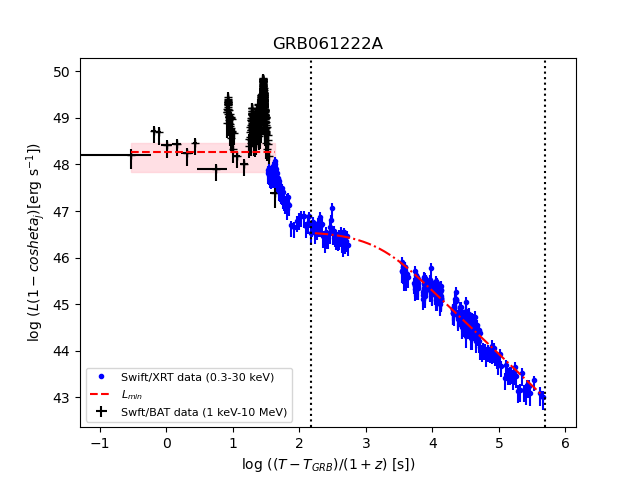}
\includegraphics[scale=0.365]{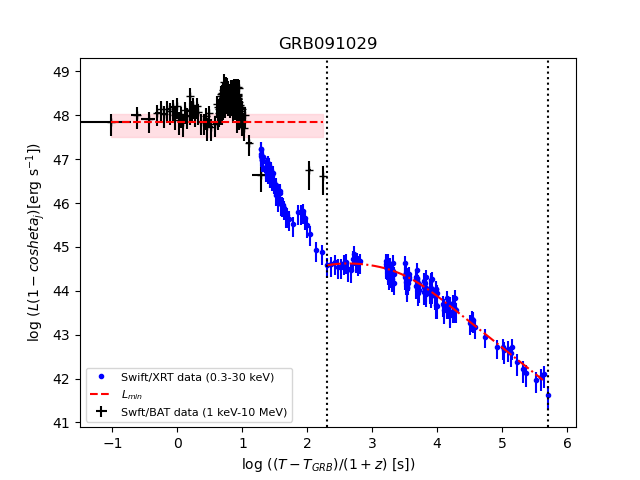}
\includegraphics[scale=0.365]{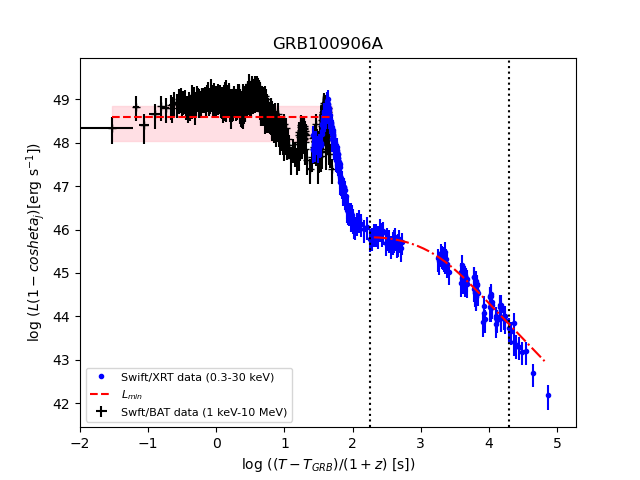}
\includegraphics[scale=0.365]{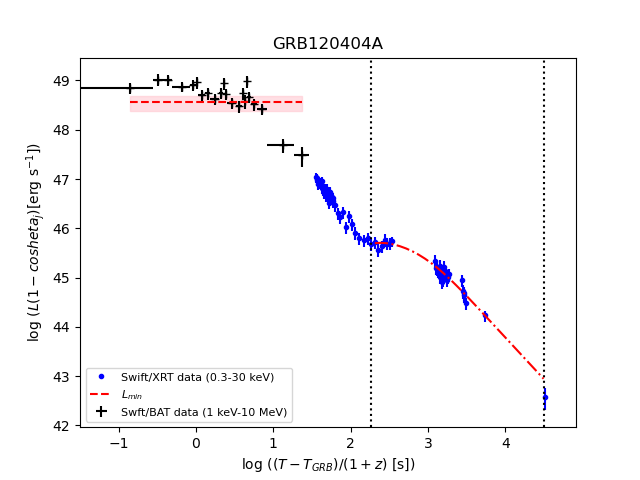}
\includegraphics[scale=0.365]{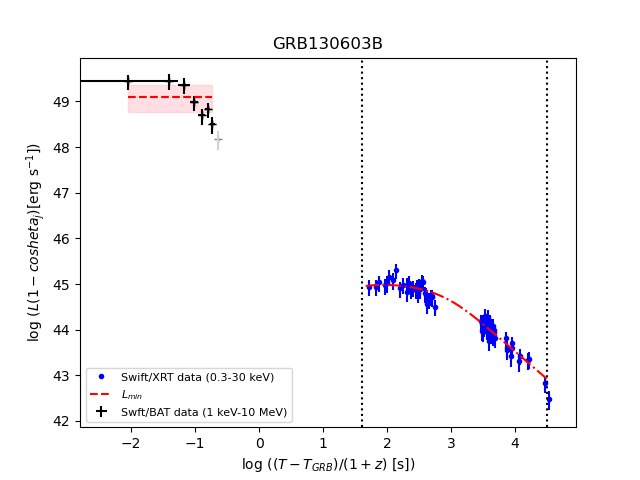}
\includegraphics[scale=0.365]{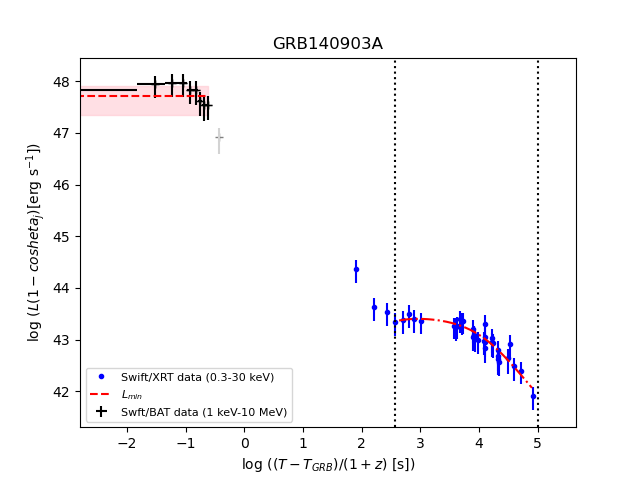}
\includegraphics[scale=0.365]{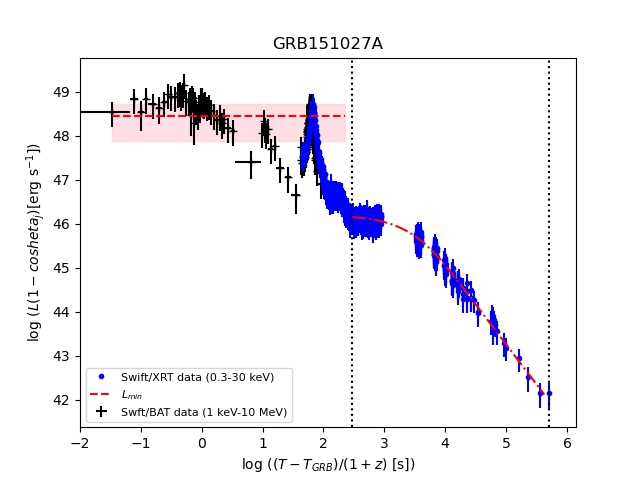}
\caption{GRB prompt and afterglow luminosity light curves in the (1 keV - 10 MeV) and (0.3-30 keV) rest frame energy range, respectively. The luminosity is corrected for jet beaming. Black vertical dotted lines indicate the temporal window considered for the magnetar model fit.~Red dash-dotted lines indicate, for each GRB, the best fit magnetar model from which we obtain the magnetic field strenght $B$ and spin period $P$ (see Tab.2). Horizontal dashed red lines mark the minimum accretion luminosity ($L_{\rm min}$) before entering the propeller regime, and the pink shading shows the uncertainty on $L_{\rm min}$.~The latter is wider than the uncertainty on $F_{\rm min}$ (Fig.~\ref{fig:batlc}), due to the added uncertainty on the beaming factor ($f_b$).}
\label{fig:xray_lc}
\end{figure*}

\subsection{Determination of $L_{\rm min}$}
We proceed to compute the 1-10$^4$ keV minimum luminosity $L_{\rm min}$ of each GRB as 
\begin{equation}
\label{eq:flusso-to-lumi}
    L_{\rm min}=K F_{\rm min}4\pi D_L^2 f_b\,,
\end{equation}
where $f_b=(1-\cos\theta_j)$ is the beaming  factor and $D_L$~the~luminosity distance.~The K-correction converts the (15-150)~keV observed fluxes to the $(1-10^4)$~keV rest-frame ones
\begin{equation}
\label{eq:kcorrection}
    K =\frac{\int_{1~{\rm keV}/(1+z)}^{10^4 {\rm keV}/(1+z)} F(E)dE}{\int_{15~{\rm keV}}^{150~{\rm keV}} F(E)dE}\,,
\end{equation}
where $F(E)$ is the prompt emission best fit spectral model\footnote{Best-fit models refer to the GRB integrated spectra rather than~the peaks.}, taken from the joint Konus-Wind~(20 keV-15 MeV) and Swift/BAT~(15-150 keV) GRB catalogues\footnote{GRB 140903 is the only event not present in these catalogues.~Its spectral model is discussed in sec.~\ref{sec:140903}.} (\citealt{tsve17, tsve21}).~Table~\ref{tab:tab0} reports the adopted spectral model and its parameters, the observed prompt peak energy $E_p$ and the isotropic equivalent energy $E_{\rm iso}$  in the $(1-10^4)/(1+z)$ keV band from the same catalogue.

The uncertainty on $L_{\rm min}$ includes the one on the jet opening angle as well as the 
 uncertainty inherent in the adopted value of $F_{\rm min}$, and is computed as: 
\begin{equation}
    dL_{\rm min} = K 4\pi D^2_L \sqrt{[(f_b dF_{\rm min})^2 + (F_{\rm min} df_b)^2]}\, .
\end{equation}
Here we adopted $df_b \approx \sin\theta_j d\theta_j$ if $d\theta_j < 0.5~\theta_j$, otherwise $df_b=(f_{b,{\rm max}}-f_{b, {\rm min}})/2$ where $f_{b,{\rm max}}$ and $f_{b, {\rm min}}$ are the collimation factors corresponding to $\theta_j+d\theta_j$ and $\theta_j-d\theta_j$.~When $\theta_j$ is provided in the literature without an error, we assume a conservative $d\theta_j=0.5\theta_j$ and apply the latter formula.~In general, we find that both the uncertainties on $\theta_j$ and on 
$F_{\rm min}$ contribute significantly to the resulting uncertainties~in~$L_{\rm min}$. 
\subsection{Parameter estimation for magnetars}
\label{sec:magnetarspindown}

To estimate the NS $B$-field and spin period we fit the~X-ray\\ afterglow plateaus with the model of energy injection from a spinning down magnetar developed by \citet{dallosso11} and further generalised in \citet{stratta18}.~In the  model, the plateau luminosity ($L_p$) and duration ($T_p$) are proxies of the same quantities in the magnetar spindown law, which in turn are determined by $B$ and $P$ (e.g. Fig.~1 in \citealt{dallosso11}).~In particular, a straightforward interpretation of fit results is enabled by use of the standard approximations (e.g.~\citealt{dallo21}),
\begin{eqnarray}
\label{eq:dedt}
T_p & \sim & 6.8 \times 10^4  \left(\displaystyle \frac{P_{\rm ms}}{B_{14}}\right)^2~{\rm s}\nonumber \\
L_p(t) & \sim &  \epsilon_{\rm sd} \displaystyle \frac{E_{\rm spin, i}} {T_p} \left(1+ \displaystyle \frac{t}{T_p}\right)^{-2} \, ,
\end{eqnarray}
where $B_{14}$ is the NS dipole magnetic field in units of $10^{14}$~G, $P_{\rm ms}$ its initial spin period in milliseconds, $E_{\rm spin,i}$ its initial spin energy, and we added $\epsilon_{\rm sd}$, the conversion efficiency from the NS spindown power to afterglow X-ray emission. 

The plateau X-ray luminosity is computed as $L_{p,X}(t)= K^{\prime} F_X(t) 4\pi D^2_L f_b$, where the X-ray flux $F_X(t)$ is taken from the Swift/XRT Burst Analyzer repository (corrected for absorption, see \citealt{Evans10}).~Here the K-correction converts from the observed (0.3-10)~keV band to the rest frame (0.3-30) keV band.~The beaming factor $f_b$ is the same as that adopted for the prompt emission in eq.~\ref{eq:flusso-to-lumi}~(see Tab.~2).  

The magnetar luminosity is assumed 
to be released isotropically during the plateau, as typically done in the literature (e.g. \citealt{Rowlinson13}; \citealt{pirol19}).~Consequently,~in  order to calculate the rate of energy injection in the external shock, the luminosity $L_{\rm  sd}$ was reduced by the fraction~of solid angle intercepted by the GRB jet, {\it i.e.}~the beaming factor $f_b$.~Fig.~\ref{fig:xray_lc} shows the beaming-corrected afterglow light curves of the 9 GRBs in our sample in the 0.3-30 keV rest-frame energy range.~Superposed are the best fit energy injection models. 
~The best fit $B$ and $P$ values are reported in~Tab.~\ref{tab:tab1}.
\begin{table*}[t]
\footnotesize
\centering
\begin{tabular}{|c|c|c|c|c|c|c|c|c|c|}
\hline
GRB & $\theta_j$ & $F_{{\rm min},15-150~ {\rm keV}}$ & $L_{{\rm min},1-10^4~ {\rm keV}}$ & $P$ & $B$ & $\epsilon$ & $L_{p,X, {\rm 0.3-30 ~ keV}}$ & $\epsilon_{\rm sd}$ \\ 
 ~  &  [rad]      & [$10^{-7}$ cgs]  & [$10^{47}$ erg/s]    & [ms] & [$10^{14}$ G] & & [$10^{44}$ erg/s] &  \\
\hline
060614 & $ 0.18 $ & $ 0.32 ^{+ 0.3 }_{ -0.2 }$ & $ 0.6 ^{+ 0.7 }_{- 0.5 }$ & $ 30.3 \pm 0.5 $ & $ 50.3 \pm 1.9 $   &  $ 0.07 ^{+  0.10  }_{- 0.07 }$ & $0.08 \pm 0.04$ &$0.17 \pm 0.04 $ \\

060729A & $ 0.114 ^{+ 0.004 }_{- 0.004 } $ & $ 0.16 ^{+ 0.09 }_{ -0.09 }$ & $ 3.5 ^{+ 2.0 }_{- 2.0 }$ & $ 2.90 \pm 0.03 $ & $ 4.63 \pm 0.08 $   &  $ 0.23 ^{+  0.14  }_{- 0.14 }$ & $2.6 \pm 0.2$ & $0.16 \pm 0.04$ \\

061222A & $ 0.0684 ^{+ 0.014 }_{- 0.006 } $ & $ 0.08 ^{+ 0.04 }_{ -0.04 }$ & $ 18.8 ^{+ 10.0 }_{- 12.0 }$ & $ 0.87 \pm 0.02 $ & $ 7.01 \pm 0.35 $   &  $ 0.03 ^{+  0.02  }_{- 0.02 }$ & $500 \pm 200$  & $0.27 \pm 0.07$\\

091029 & $ 0.022 ^{+ 0.016 }_{- 0.016 } $ & $ 0.13 ^{+ 0.03 }_{ -0.03 }$ & $ 7.0 ^{+ 3.7 }_{- 3.7 }$ & $ 1.54 \pm 0.09 $ & $ 9.5 \pm 0.5 $   &  $ 0.024 ^{+  0.01  }_{- 0.01 }$ &$6.0\pm3.8$ & $0.17 \pm 0.06$  \\

100906A & $ 0.05 $ & $ 0.6 ^{+ 0.3 }_{ -0.3 }$ & $ 38^{+ 33}_{- 27}$ & $ 1.72 \pm 0.07 $ & $ 16.2 \pm 0.9 $   &  $ 0.06 ^{+  0.05  }_{- 0.05 }$ &$110 \pm 70$ &$0.34 \pm 0.09$ \\

120404A & $ 0.0541 ^{+ 0.005 }_{- 0.005 } $ & $ 0.17 ^{+ 0.05 }_{ -0.05 }$ & $ 36 ^{+ 13 }_{- 13 }$ & $ 2.54 \pm 0.23 $ & $ 34.1 \pm 1.8 $   &  $ 0.03 ^{+  0.01  }_{- 0.01 }$ &$43\pm9$ &$0.12 \pm 0.05$ \\

130603B & $ 0.11 ^{+ 0.03 }_{- 0.09 } $ & $ 3.8 ^{+ 3.0 }_{ -1.5 }$ & $ 126 ^{+ 110}_{- 67}$ & $ 13.3 \pm 1.1 $ & $ 153 \pm 11 $   &  $ 0.26 ^{+  0.25}_{- 0.16 }$ & $8.7 \pm 3.6$ &$0.22 \pm 0.09$  \\

140903A & $ 0.07 ^{+ 0.09 }_{- 0.03 } $ & $ 1.5 ^{+ 0.5 }_{ -0.5 }$ & $ 5.1 ^{+ 2.9 }_{- 2.9 }$ & $ 13.7 \pm 1.5 $ & $ 39 \pm 5$   &  $ 0.17 ^{+  0.13 }_{- 0.13}$ &$0.3\pm0.2$ & $0.32\pm 0.16$ \\

151027A & $ 0.11 $ & $ 0.66 ^{+ 0.44}_{-0.34}$ & $ 28^{+ 26}_{-20}$ & $ 2.00 \pm 0.01 $ & $ 12.7 \pm 0.2 $   &  $ 0.10 ^{+ 0.10}_{-0.08}$ & $150 \pm 90$ & $0.29 \pm 0.04$\\

\hline
\end{tabular}
\caption{Selected sample of 9 GRBs analyzed in this work.~The magnetar energy output is assumed to be emitted isotropically and the NS angular momentum is assumed to be orthogonal to the magnetic axis.~The second column gives the jet opening angle, while the third and fourth columns quote the minimum flux (Swift/BAT) and corresponding (bolometric) luminosity in the prompt phase before the NS enters the propeller regime. $P$ and $B$ are the magnetar spin period and dipole $B$-field strength obtained by fitting the X-ray afterglow light curves, and $\epsilon$ the conversion efficiency from gravitational potential energy to $\gamma$-rays.~The eighth column quotes the (beaming-corrected) plateau luminosity in the earliest time bin, while the last 
column reports the inferred conversion efficiency from NS spindown power to plateau luminosity.}
\label{tab:tab1}
\end{table*}
\begin{figure*}[ht]
\centering
\includegraphics[scale=0.6]{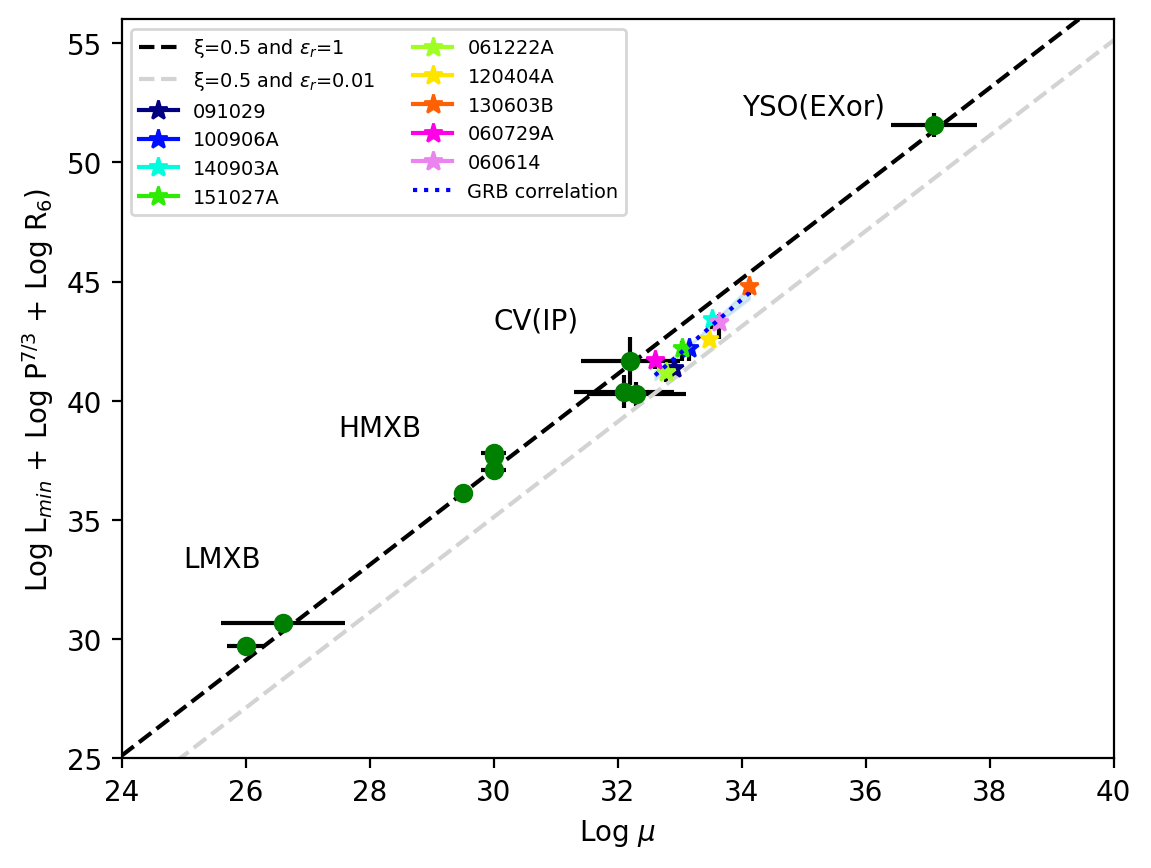}
\includegraphics[scale=0.6]{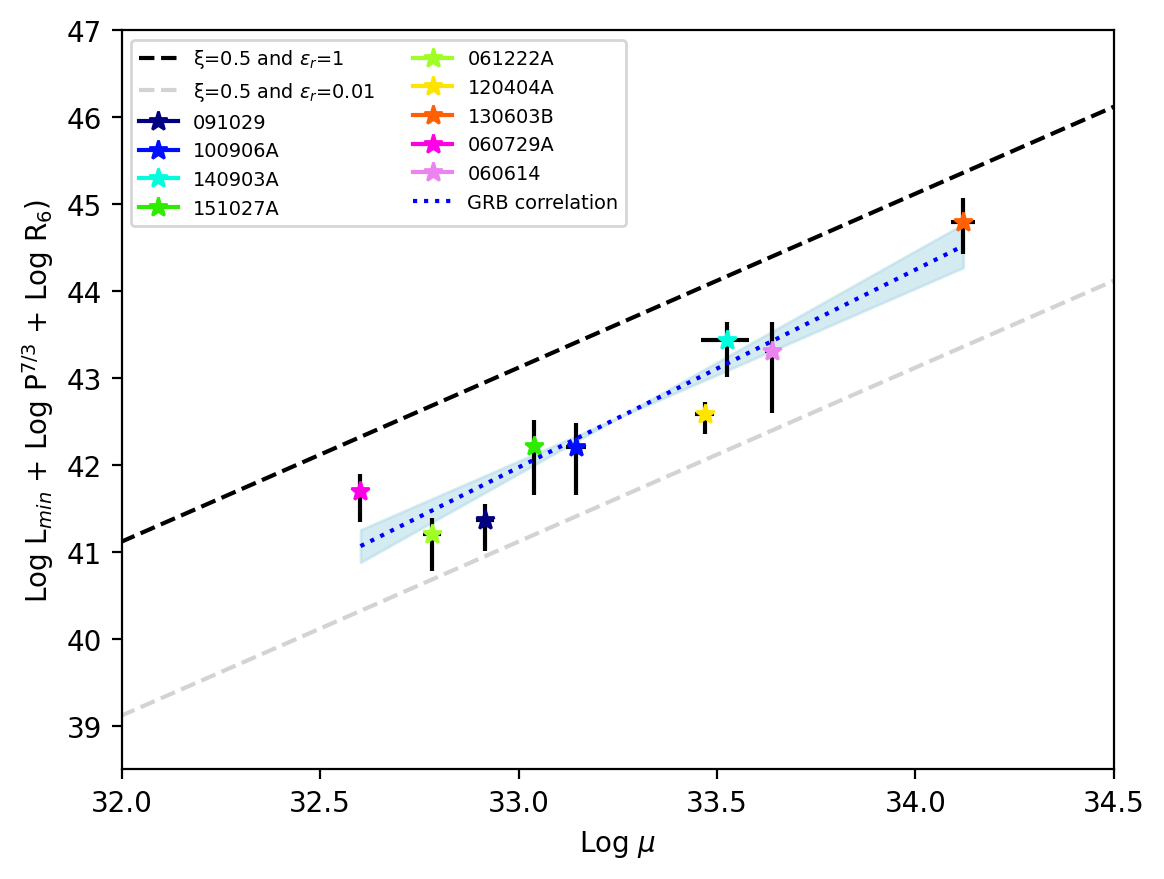}
\caption{{\em Left:}~The Universal relation for two different radiative efficiencies, $\epsilon=1$ (dashed thick line) and $\epsilon=0.01$ (dashed light-gray line; see Eq.~\ref{eq:Bell-relation}).~Green dots indicate different classes of accreting compact objects studied by \citet{campana18}.~Coloured stars are the GRBs analyzed in this work and for which an isotropically radiating  magnetar central engine is assumed.~The blue dotted line is the best fit correlation for the 9 GRBs, with a 
Pearsons correlation coefficient 
$= 0.948$ corresponding to a two-tailed $p$-value for testing a non-correlation of $9.7\times10^{-5}$.~{\em Right:}~Close-up of the left panel, highlighting the GRB population alone.~The shaded area around the correlation blue dotted line shows the uncertainty in the fit parameters.}
\label{fig:universe_mag90}
\end{figure*}
\section{Results: connecting theory to the data}
\label{sec:results}
With the derived values of $L_{\rm min}$, $B$ and $P$, reported in Tab.~\ref{tab:tab1}, each GRB is plotted in the $L_{\rm min} P^{7/3} R_6$ vs.~$\mu$ plane, for a NS radius of 12~km, in order to test their consistency with the universal relation (Eq.~\ref{eq:Bell-relation}). 
\subsection{Universal relation}
We find a strong correlation for the GRBs in our sample in the $\log_{10}(L_{\rm min} P^{7/3} R_6)~{\rm vs}$ $\log_{10} \mu_{30}$ plane, with a Pearson correlation coefficient $r$ = 0.948 and two-tailed $p$-value 
$=9.7\times10^{-5}$. If we discard the two short GRBs (GRB~130603B and GRB~140903A), for which $L_{\rm min}$ is more uncertain, the correlation remains significant, $r=0.891$ and $p$-value~=~0.007. The correlation has the form\footnote{It is $y=(1.8 \pm 0.3)x + (36.2 \pm 1.2)$ if we remove the two short GRBs.} 
~$y=(2.3\pm 0.3)x +(35\pm1)$~in~the plane of Fig.~\ref{fig:universe_mag90}; the slope is consistent, within the uncertainties, with the one expected from theory (Eq.~\ref{eq:Bell-relation}) and observationally confirmed by \citet{campana18} for other types of accretion-powered sources hosting a magnetic stellar~object, $y =2x + (37.12\pm 0.20)$ 
with $\xi=0.5$ and $\epsilon=1$. Compared to the latter, the GRBs in our sample require a lower normalization, which is readily interpreted in terms of a reduced radiative efficiency of their central engines (Eq.~\ref{eq:Bell-relation}), as predicted by theory.~Indeed, as discussed in sec.~{\ref{sec:peculiar},~the~efficiency of conversion of gravitational potential energy to prompt radiation ($\epsilon$) is expected to be low in GRB discs.~From Eq.~\ref{eq:Bell-relation} we estimate for each GRB the value of\footnote{In eq.~\ref{eq:estimate-epsilon} we omit for simplicity a dependence on $M_{1.4}^{2/3}$.} $\epsilon$ implied by its position in the plane of Fig. \ref{fig:universe_mag90}, \\ 

\be
\label{eq:estimate-epsilon}
\log_{10} \epsilon =  \log_{10} y - 2 \log_{10} \mu_{30} - \displaystyle \frac{7}{2} \log_{10}  \displaystyle \left(\frac{\xi}{0.5}\right) - 37.15
\ee
and derive typical values of $\epsilon \sim (0.03-0.26)$, as reported~in Tab.~\ref{tab:tab1}.~The corresponding conversion efficiencies from rest-mass to prompt radiation, 
$\eta \sim \left(0.5-4 \right) \times 10^{-2}M_{1.4}R_6^{-1}$, 
 are remarkably consistent with the range estimated from analysis of GRB prompt and afterglow lightcurves (e.g. \citealt{zhang07}), with the two short GRBs in the sample requiring the largest efficiencies. 

Factors like the mass and radius of each individual NS may contribute to the scatter of the correlation; a more crucial role in eqs.~(\ref{eq:Bell-relation})~and~(\ref{eq:estimate-epsilon}) may be played by the $\xi$-value.~\cite{campana18} showed that the data point to $\xi \approx~0.5$ in their composite sample of sub-Eddington sources.~We note that, if $\xi \approx 1$, as some authors suggest for very high accretion rates (e.g.~\citealt{andersson05, kulka13}), then  the estimated $\epsilon$ in Eq.~\ref{eq:estimate-epsilon} could be further reduced by up to a factor $\lesssim 10$. 
\subsection{Testing the energy source: disc accretion and NS spin}
We further  corroborate this scenario by showing that, for each GRB, the accretion luminosity $L_{\rm min}$ in the prompt phase {and the afterglow luminosity at the onset of the plateau (as measured in its first temporal bin)} satisfy the theoretically expected value (eq.~\ref{eq:ratio}), once the accretion radiative efficiency ($\epsilon$) for that GRB is factored in.~To this aim we calculate, for each GRB in~our~ sample,~the ratio $  L_{\rm min}/L_{p, {\rm iso}} = L_{\rm min}/(\epsilon_{\rm sd} L_{\rm sd}) = \kappa / \epsilon_{\rm sd}$ using the $L_{\rm min}$-values from the previous section and the isotropic-equivalent plateau luminosity, $L_{p, {\rm iso}}  = \epsilon_{\rm sd} L_{\rm sd}$, as a proxy for the NS spindown power.~As in eq. \ref{eq:dedt}, we include the efficiency $\epsilon_{\rm sd}$ to allow for non-perfect conversion of spindown power into plateau emission.~We then equate the observationally-determined $L_{\rm min}/L_{p, {\rm iso}}$ to its expected value (see eq. \ref{eq:ratio}), $\kappa/\epsilon_{\rm sd} \approx 10^5~\epsilon/\epsilon_{\rm sd} P^{5/3} (\xi/0.5)^{7/2}$ (for $R=12$~km and $M=1.4 M_{\odot}$) and derive the ratio $\epsilon/\epsilon_{\rm sd}$.~For each GRB, Tab.~2 reports the resulting value of $\epsilon_{\rm sd}$ adopting the best-fit spin period from the plateau and the $\epsilon$ derived from Eq.~\ref{eq:estimate-epsilon}.
We obtain typical values of $\epsilon_{\rm sd} \sim 0.12-0.34$, remarkably similar for all the GRBs in our sample, confirming within the uncertainties the self-consistency of the proposed scenario.

Note that, if the spindown were enhanced by a factor $\beta$ relative to the ideal magnetic dipole formula of Eq.~\ref{eq:spindown} (e.g., \citealt{parfrey16, metzger18}), then a somewhat lower $\epsilon_{\rm sd}/\beta$ would result.~In addition, our results would~be~weakly~affected even if the~magnetar wind, responsible for energy injection, were mildly beamed e.g.~by a factor $f_w > 0.14$ ({\it i.e.} $\theta_w > 30^\circ$) along the jet axis.~At a fixed jet half-opening $\theta_j$, beaming of the magnetar wind would imply that a larger fraction of its spindown power is intercepted by the jet, relative to an isotropic wind.~Thus, the required luminosity of the magnetar would also be lower by a factor $f_w$.~We have verified that, as a result of this change, each point in Fig.~\ref{fig:universe_mag90} will  shift towards regions of slightly lower radiative efficiency.~In particular, $\epsilon$ would be reduced by just $f^{1/6}_w$, while $\epsilon_{\rm sd}$ would increase by $f_w^{-2/3}$, relative to 
our fiducial values in Tab.~\ref{tab:tab1}.

\subsection{Propeller, ejector and NS dynamics}
\label{sec:self}
\begin{table*}
\footnotesize
\centering
\begin{tabular}{|c|c|c|c|c|c|c|c|c|c|c|}
\hline
GRB & $\dot{m}_{\rm min}$ & T$_{\rm min}$ & $r_m(T_{\rm min})/R$ & $T_{\rm pl}$ & $r_L(T_{\rm pl})/R$ &  
$r_m/r_L (T_{\rm pl})$ & $P(T_{\rm min})$ & $P(T_{\rm pl})$ & $\Delta M_{\rm acc}$ \\
    &  ($10^{-4}~M_\odot$/s)   & (s)   & &(s) & & & (ms) &  (ms) & $(10^{-3} M_\odot)$\\   
\hline
060614 
&  0.15 & 85 & 8.5 & $< 4500$ & 120.4 & $< 0.75$  & 15.1 &30.3 &  $1.6$\\
060729A 
& 0.05 & 90 & 2.82 & 500 & 11.5 & 0.57 & 2.88 & 2.9 & 0.5 \\ 
061222A 
&  2.2 & 35 & 1.24 & 100 & 3.5& 0.6 & 0.84 & 0.87 & 5.8 \\
091029 &  1.0 & 10.5 
& 1.83 & 130 & 6.1 &  0.99 & 1.50 &1.54 & 1.3\\
100906A* &  2.6 & 35 
& 1.86 & 130 & 6.8 & 0.52 & 1.54 & 1.72 & 7.9\\
120404A* 
&  5.6 & 6.5 & 2.3 & 110 & 10.1 & 0.9 & 2.13 & 2.54 & 4.6  \\
130603B & 1.7 & 0.09 & 7.7 & < 30 & 52.9 & < 2.3 & 12.9 & 13.3 & 0.02\\
140903A &  0.1 & 0.15 & 7.95 & < 100 & 54.5 & <3.2 & 13.6 & 13.7 & 0.005\\
151027A  & 1.3 & 60 
& 2.01 & 350 & 7.95 & 0.6 & 1.73 & 2.00 & 8.1\\
\hline
\end{tabular}
\caption{Additional physical parameters of magnetar central engines derived in our GRB sample.~The mass accretion rate $\dot{m}_{\rm min}$ at $T_{\rm min}$ is calculated using $\epsilon$ from Tab.~\ref{tab:tab1}.~The times $T_{\rm min}$ and $T_{\rm pl}$ are obtained directly from the prompt lightcurve.~The magnetospheric radius $r_m$ (in units of the NS radius, $R$) at $T_{\rm min}$ uses $\dot{m}_{\rm min}$ and the B-field derived from fits to the X-ray plateau.~The light-cylinder radius $r_L$ at the start time of the plateau ($T_{\rm pl}$) uses the best-fit spin period for the X-ray plateau, {\it i.e.} $P(T_{\rm pl})$.~The value of $r_m(T_{\rm pl)}$ is calculated from $r_m(T_{\rm min})$ using $\dot{m}_{\rm min}$ and the scaling $\dot{m} \propto t^{-5/3}$. 
$P(T_{\rm min})$ is the NS spin period at $T_{\rm min}$, calculated with our model for the propeller-induced spindown.~The last column features the total amount of mass that went through the disc during the propeller phase, {\it i.e.} between $T_{\rm min}$ and $T_{\rm pl}$.}
\label{tab:3}
\end{table*}
At time $T_{\rm min}$ the prompt luminosity drops below $L_{\rm min}$ and the NS enters the propeller regime.~Here the interaction of the magnetosphere with the (slower) material in the Keplerian disc causes a phase of enhanced spindown with respect to eq.~\ref{eq:spindown}.~This phase will end only when the NS switches to the ejector regime, in which the disc is expelled from the magnetosphere\footnote{ 
Matter flung out by the propeller, if not unbound, may return to the disc and drift inward on viscous timescales (e.g.~\citealt{Li21}),~by which~time the ejector sweeps away the residual disk together with any recycled material, thus preventing significant effects on source luminosity.} (as its inner edge approaches the light cylinder) and the NS 
spins down due to magnetic dipole radiation (eq.~\ref{eq:spindown}).
As a result the spin period $P$ estimated from the plateau may be longer than $P(T_{\rm min})$, {\it i.e.} the spin period at time $T_{\rm min}$. Moreover, 
the luminosity evolution from $T_{\rm min}$ through $T_{\rm pl}$, the plateau start time, will track the power released in the disc down to $r_m$. We check here the consistency of observed GRB light curves with the above expectations.~A more systematic study of light curve shapes as a function of model parameters is~deferred for future investigation. 

Our first step is to check the impact of propeller-induced spin down on the NS rotation from $T_{\rm min}$ to $T_{\rm pl}$.~For each~GRB we estimate $\dot{m}_{\rm min} = L_{\rm min} R/(GM\epsilon)$ at~time~$T_{\rm min}$, and the corresponding $r_m (T_{\rm min})$.~However, we do not use here the previously calculated radiative efficiency since, as eq.~{\ref{eq:estimate-epsilon} shows, $\epsilon \sim L_{\rm min} R P(T_{\rm min})^{7/3}/(\mu^2 \xi^{7/2})$;~thus $\epsilon$, $\dot{m}_{\rm min}$ and $r_m(T_{\rm min})$, will all depend on the unknown  $P(T_{\rm min})$.

For the mass inflow rate in the propeller phase we adopt the standard scaling $\dot{m} \propto t^{-5/3}$ for fallback (e.g.~\citealt{rees88}; \citealt{macfadyen99}).~The correspoding accretion power released in the disc is $L_{\rm prop}(t) = \epsilon GM \dot{m}(t)/2r_m(t)$.~Writing $\dot{m}(t) = \dot{m}_{\rm min}(t/T_{\rm min})^{-5/3}$, and since $r_m \propto \dot{m}^{-2/7}$~(eq.~\ref{eq:magnetosph}), then
\be
\label{eq:prop}
L_{{\rm prop},47}(t) \approx \frac{M^{8/7}_{1.4} \epsilon_{-2}\dot{m}^{9/7}_{{\rm min}, -4}}{(\xi/0.5)\mu_{33}^{4/7}} \left(t/T_{\rm min}\right)^{-15/7} \, ,
\ee
with $\dot{m}_{{\rm min}, -4} = \dot{m}_{\rm min}/(10^{-4} M_{\odot}/{\rm s})$~and~$\mu_{33} = \mu / (10^{33}{\rm G cm}^3)$.

The maximum rate of angular momentum transfer from the NS to disc material is (e.g. \citealt{piro11, parfrey16, metzger18})  $\dot{J}_{\rm prop}(t)= \dot{m}(t) \omega(t) r_m(t)^2 = -I \dot{\omega}(t)$, from which we obtain the evolution of the NS spin frequency,
\be
\label{eq:omega-prop}
\omega(t) = \omega (T_{\rm min}) \exp{\left[-\displaystyle \frac{\int_{T_{\rm min}}^t \dot{m}(t^\prime) r^2_m(t^\prime) dt^\prime}{I} \right]} \, .
\ee
After some manipulation the integrand in eq.~{\ref{eq:omega-prop} becomes
\be
\dot{m}(t) r^2_m(t) = \displaystyle \frac{\pi \sqrt{2} \xi^{7/2} \mu^2}{GM P(T_{\rm min})} \left(t/T_{\rm min}\right)^{-5/7} \, ,
\ee
which is solved for $P(T_{\rm min})$ once we determine the start time of the plateau ($T_{\rm pl}$) and  impose that $P(T_{\rm pl})$ equals the best-fit value $P$ from Tab.~\ref{tab:tab1}.~Thus, for each GRB in our sample we first estimated $T_{\rm min}$ and $T_{\rm pl}$ from the light curves, and then solved eq.~\ref{eq:omega-prop} for $P(T_{\rm min})$.~We show in Tab.~\ref{tab:3} the resulting $P(T_{\rm min})$ for each GRB in our sample, along with the values of $T_{\rm min}$, $T_{\rm pl}$, and the implied $\dot{m}_{\rm min}$.~In all but one case the spin change is small or negligible, compared to the fit error, and the same holds for the $\epsilon$-value.~Only GRB 060614 requires a $\sim$ 2 times faster spin at $T_{\rm min}$ owing to its longer initial spin period which leads to a prolonged propeller phase. 

This result fully supports the robustness of the universal relation, and its interpretation discussed in previous sections\footnote{We note in passing that our conclusion agrees with previous models showing that the millisecond magnetar spins down substantially only for propeller durations longer than hundreds of seconds (e.g. \citealt{piro11}).}.\\Moreover, the proposed scenario provides us with a simple self-contained model for GRB lightcurves from the onset of the prompt steep decay to the end of the afterglow plateau, which depends only on three parameters: $L_{\rm min}, B$ and $P$.~As an example, we show here the application of this model to the light curve of GRB 091029: having determined $B$ and $P$ from the afterglow plateau, and $L_{\rm min}$ from the prompt emission, we obtain $\epsilon$ through eq.~\ref{eq:estimate-epsilon}, which then fixes $\dot{m}_{\rm min}$, hence the propeller luminosity evolution and the propeller end time, {\it i.e.} the onset of the ejector phase.~Two conclusions~are apparent: (i) $L_{\rm prop} (t) \propto t^{-15/7}$~is well consistent with the steep decay observed at the end of the prompt phase and (ii) the onset of the plateau occurs at the time when $r_m \approx r_L$.  

Similar results are obtained for all GRBs in our sample.~For each of them, Tab.~\ref{tab:3} shows the location of $r_m$ at $T_{\rm min}$ and at the start of the plateau ($T_{\rm pl}$): GRB 091029 and GRB 120404A match very well our model prediction, and the two short GRBs 130603B and 140903A are well consistent with it, given the time coverage of the available data.~In the remaining 5 GRBs the plateau appears to start when $r_m$, though expanded by a large factor, is still $\sim (0.5-0.7) r_L$.~In all of these 5 events, the prompt emission displays two or more peaks well separated in time, suggesting distinct accretion episodes with different start times relative to the GRB trigger.~As shown, e.g., in \cite{dallosso17}, such a temporal offset offers a simple interpretation for the apparent mismatch and may help reconcile these 5 GRBs with the basic picture proposed here.~In particular, in order for $r_m \sim (0.9-1) r_L$ at the start of the plateau in these GRBs, an offset time $\sim (0.5-0.7) T_{\rm min}$ is typically required.~A detailed modelling of this and other effects that may contribute to the light curve shape is under way and will be presented in a forthcoming paper.  
~\\
\begin{figure}
\centering
\includegraphics[width=0.52\textwidth]
{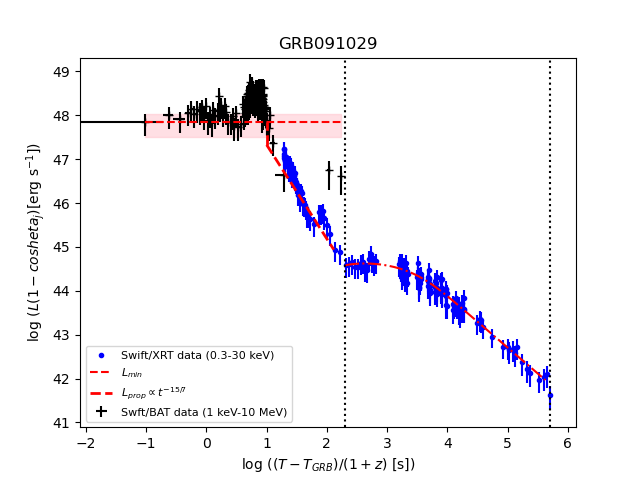}
\caption{GRB 091029 light curve from the prompt phase to~the end of the plateau.~The red dot-dashed curve shows our model prediction for the minimum accretion luminosity, $L_{\rm min}$, the propeller luminosity, the switch on time for the ejector phase ($r_m = r_L$) and the plateau luminosity.~Note that this is not a fit, but the curve expected given ($L_{\rm min}, B, P$).~The early part of the flux decay is somewhat steeper than the propeller slope, likely reflecting the standard high-latitude emission from a relativistic jet. Later it settles to the $\sim t^{-15/7}$ expected from propeller.}
\label{fig:propeller-ejector}
\end{figure}
~\\
~\\
\section{Summary \& Conclusions}
\label{sec:discuss}
Several GRBs, characterized by a plateau in their X-ray afterglow light curves, have been modeled under the assumption of energy injection by a magnetar.~In this work we have gone one step further in testing the presence of a magnetar central engine, by singling out a physical mechanism for accretion onto magnetic neutron stars - the transition to propeller - that has long been verified for NS in X-ray binaries and more recently validated for other classes of accreting stellar-mass objects \citep{campana18}.~Under the assumption that the prompt emission of GRBs is powered by a phase of mass accretion towards~a fast-spinning magnetar, we  identified signatures of propeller transitions in the prompt light curves of GRBs.~Based on such signatures, we estimated the minimum accretion  luminosity, $L_{\rm min}$, in a sample of 9 GRBs for which sufficient data are available.

We then estimated the dipolar magnetic field and the~spin period of the NSs through fits to the X-ray afterglow~plateaus. By combining these {\it independently} determined parameters according to Eq.~\ref{eq:Bell-relation}, we showed that the GRBs in our sample follow~the same universal relation between $L_{\rm min}$, $P$ and $B$ as other accreting, magnetic stellar-mass objects, albeit at a lower radiative efficiency $\epsilon \sim 10^{-2}-10^{-1}$.~The latter is also in good agreement with independent observational constraints (e.g. \citealt{zhang07}) and theoretical expectations.

Moreover, {by simply comparing the isotropic-equivalent luminosity of the X-ray plateau with that predicted by the NS spindown model, we have estimated a conversion efficiency of spindown power to afterglow emission $\sim 0.12-0.34$, remarkably consistent throughout our sample.~This illustrates~the diagnostic power of the ratio $L_{\rm min}/L_{p,X}$  (Eq.~\ref{eq:ratio}), which depends only on the NS spin period and the ratio of the radiative efficiencies $\epsilon / \epsilon_{\rm sd}$ in each  source.~Our results demonstrate that a magnetar central engine can account at once for:~(i)~the prompt luminosity at the onset of the steep decay, the plateau duration and its luminosity in each individual GRB and~(ii)~a correlation among these properties in the GRB population, while simultaneously deriving radiative efficiencies of both accretion and spindown in agreement with standard values and (iii) the slope of the prompt steep decay, as well as the onset time of the plateau. The magnetar scenario we discussed - where accretion powers the GRB prompt emission and, once accretion subsides, the NS spindown powers the afterglow plateau - is amenable to further development aimed at fitting the different stages of GRB lightcurves with a minimal set of model parameters. This will be the subject of future work.

\begin{acknowledgments}
We thank the anonymous referee for very helpful and insightful comments.~This project has received funding from the European Union’s Horizon2020 research and innovation programme under the Marie Skłodowska-Curie (grant agreement No. 754496).~R.P.~ kindly acknowledges support by NSF award AST-2006839. G.S. acknowledges the support by the State of Hesse within the Research Cluster ELEMENTS (Project ID 500/10.006). 
L.S. acknowledges financial contributions from ASI-INAF agreements 2017-14-H.O and  I/037/12/0; from “iPeska” research grant (P.I. Andrea Possenti) funded under the INAF call PRIN-SKA/CTA (resolution 70/2016), and from PRIN-INAF 2019 no. 15.~This work made use of data supplied by the UK Swift Science Data Centre at the University of Leicester. 
\end{acknowledgments}
\bibliographystyle{aasjournal}
\bibliography{dallosso-GRBs-NS.bbl}
\end{document}